\documentclass[11pt,a4paper]{article}

\usepackage[a4paper,margin=1in,footskip=0.25in]{geometry}
\usepackage{natbib}
\usepackage{hyperref}
\usepackage{url}
\usepackage{amsmath}
\usepackage{amssymb}
\usepackage{slashed}
\usepackage{morefloats}
\usepackage{authblk}
\usepackage{graphicx} 
\newcommand{\nnl}{\nonumber \\}

\newcommand{\beq}{\begin{equation}} 
\newcommand{\eeq}{\end{equation}} 
\newcommand{\ba}{\begin{array}}  
\newcommand{\ea}{\end{array}} 
\newcommand{\bea}{\begin{eqnarray}}  
\newcommand{\eea}{\end{eqnarray} }  
\newcommand{\be}{\begin{eqnarray}}  
\newcommand{\ee}{\end{eqnarray} }  
\newcommand{\bal}{\begin{align}}
\newcommand{\eal}{\end{align}}   
\newcommand{\bi}{\begin{itemize}}  
\newcommand{\ei}{\end{itemize}}  
\newcommand{\ben}{\begin{enumerate}}  
\newcommand{\een}{\end{enumerate}}  
\newcommand{\bc}{\begin{center}}
\newcommand{\ec}{\end{center}} 
\newcommand{\bt}{\begin{table}}
\newcommand{\et}{\end{table}}  
\newcommand{\btb}{\begin{tabular}}
\newcommand{\etb}{\end{tabular}}

\begin{document}
%=================================================================
% Full title of the paper (Capitalized)
\title{The Standard Model theory of neutron beta decay}

% MDPI internal command: Title for citation in the left column
%\TitleCitation{The Standard Model theory of neutron beta decay}

% Author Orchid ID: enter ID or remove command
%\newcommand{\orcidauthorA}{0000-0002-3062-0118} \newcommand{\orcidauthorB}{0000-0001-9348-0557} % Add \orcidA{} behind the author's name
%\newcommand{\orcidauthorB}{0000-0000-0000-000X} % Add \orcidB{} behind the author's name

% Authors, for the paper (add full first names)
\author{Mikhail Gorchtein $^{1,2,\ddagger}$ and Chien-Yeah Seng $^{3,4,\ddagger}$}

%\longauthorlist{yes}

% MDPI internal command: Authors, for metadata in PDF
%\AuthorNames{Mikhail Gorchtein and Chien-Yeah Seng}

% MDPI internal command: Authors, for citation in the left column
%\AuthorCitation{Gorchtein, M.; Seng, C.-Y.}
% If this is a Chicago style journal: Lastname, Firstname, Firstname Lastname, and Firstname Lastname.

% Affiliations / Addresses (Add [1] after \address if there is only one affiliation.)
\affil{%
$^{1}$ \quad Institut f\"ur Kernphysik, Johannes Gutenberg-Universit\"{a}t,
	J.J. Becher-Weg 45, 55128 Mainz, Germany \\
$^{2}$ \quad PRISMA Cluster of Excellence, Johannes Gutenberg-Universit\"{a}t, Mainz, Germany \\
$^{3}$ \quad Facility for Rare Isotope Beams, Michigan State University, East Lansing, MI 48824, USA\\
$^{4}$ \quad Department of Physics, University of Washington,
	Seattle, WA 98195-1560, USA
}
\maketitle
% Contact information of the corresponding author
%\corres{Correspondence: gorshtey@uni-mainz.de (M.G.); seng@frib.msu.edu (C.-Y.S.)}

% Current address and/or shared authorship
%\firstnote{Current address: Affiliation 3.} 
%\secondnote{These authors contributed equally to this work.}
% The commands \thirdnote{} till \eighthnote{} are available for further notes

%\simplesumm{} % Simple summary

%\conference{} % An extended version of a conference paper

% Abstract (Do not insert blank lines, i.e. \\) 
\abstract{We review the status of the Standard Model theory of neutron beta decay. Particular  emphasis is put on the recent developments in the electroweak radiative corrections. Given that some existing approaches give slightly different results, we thoroughly review the origin of discrepancies, and provide our recommended value for the radiative correction to the neutron and nuclear decay rates.
The use of dispersion relation, lattice Quantum Chromodynamics and effective field theory framework allows for high-precision theory calculations at the level of $10^{-4}$, turning neutron beta decay into a powerful tool to search for new physics, complementary to high-energy collider experiments. We offer an outlook to the future improvements.    }

\section{Introduction}

Compared to other hadrons that primarily decay via weak interaction, the neutron is very long lived with the lifetime $\tau_n\approx15$ minutes. The reason lies in the small phase space available for the underlying decay $n\to p+e^-+\bar\nu_e$: the energy release $M_n-M_p-m_e\approx0.782$\,MeV is  three orders of magnitude smaller than the nucleon mass. The neutron-proton mass difference, $M_n-M_p\approx1.293$\,MeV is also smaller than the typical nuclear binding energy per nucleon which amounts to $\sim7-8$\,MeV,  making the bound neutron stable. The exact value of the neutron lifetime (together with other quantities) determines the rate of the $p-p$ fusion process that fuels the sun \cite{Bahcall:1968xb}.

A precise measurement of the neutron beta decay 
provides us with an accurate probe of  semileptonic charged-current interactions across the first generation of Standard Model (SM) fermions. The Lagrangian density responsible for the $d\to u$ quark conversion reads
\begin{align}
    {\mathcal L}_{e\nu ud} = 
- \sqrt{2}G_F{V_{ud}} 
\bar{e}  \gamma_\mu  \nu_L  \cdot \bar{u}   \gamma^\mu  (1 - \gamma_5)  d+\mathrm{h.\,c.},
\label{eq:Lenuud}
\end{align}
with $\nu_L=\frac{1}{2}(1-\gamma_5)\,\nu$ the left-handed neutrino field, $G_F=1.166\,378\,8(6)\times10^{-5}$~{GeV}$^{-2}$ the Fermi constant, and $V_{ud}$ the top-left corner element of the Cabibbo-Kobayashi-Maskawa (CKM) quark mixing matrix \cite{Cabibbo:1963yz,Kobayashi:1973fv}. With the Fermi constant known very precisely from muon lifetime measurements, $V_{ud}$ is the fundamental parameter of the SM that is of primary interest in the studies of the neutron decay. 

At low energies, the quarks are bound into hadrons, and their electroweak interactions in Eq.\eqref{eq:Lenuud} are embedded into the Lagrangian density at the nucleon level. Dropping for the moment the small nucleon recoil, the latter reads
\begin{align}
    {\mathcal L}_{e\nu pn} = 
- \sqrt{2}G_F{V_{ud}} 
\bar{e}  \gamma_\mu  \nu_L  \cdot \bar{p}   \gamma^\mu  (g_V +g_A\gamma_5)  n+\mathrm{h.\,c.}
\label{eq:Lpn_leading}
\end{align}
The conservation of the vector current ensures that $g_V=1$ (modulo tiny corrections that are quadratic in $m_d-m_u$), but the axial coupling is not protected against strong renormalization. The ratio of the two, $\lambda\equiv g_A/g_V$ has to be known sufficiently precisely to allow for an extraction of 
$|V_{ud}|$ from neutron decay~\cite{ParticleDataGroup:2022pth}:
%\begin{equation}    
%|V_{ud}|^2_n=
%\frac{4903.2(1.0)~\text{s}}{\tau_n(1+3\lambda^2)}~.\label{eq:Vudn}
%\end{equation}
\begin{equation}    
|V_{ud}|^2_n\propto
\frac{1}{\tau_n(1+3\lambda^2)}~.\label{eq:Vudn}
\end{equation}
At present, the best precision is warranted if the measurement of the lifetime is supplemented by that of $\lambda$ via one of the correlation coefficients introduced in Section \ref{sec:diffrate}. 
%
%The PDG 2022 averages read~\cite{ParticleDataGroup:2022pth} $\tau_n^\mathrm{av}=879.4(8)$\,s and $g_A^\mathrm{av}=1.2756(13)$, while the single most precise measurements are $\tau_n^\mathrm{av}=877.75(43)$\,s~\cite{}and $g_A^\mathrm{av}=1.2764(5)$

\begin{figure}[h]
    \centering
    \includegraphics[width=0.8\columnwidth]{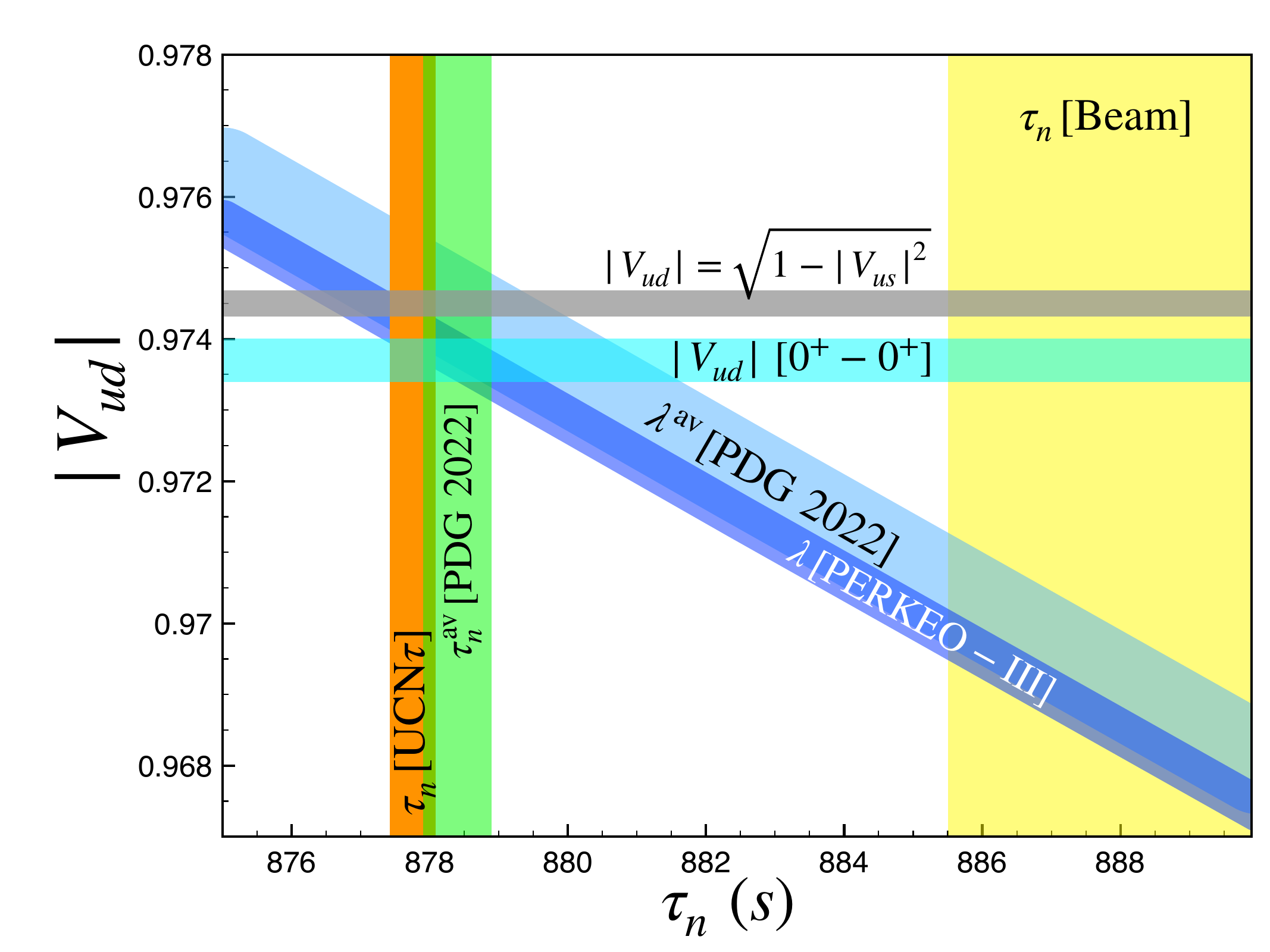}
    \caption{Experimental status of $V_{ud}$,  $\tau_n$ and $\lambda$ as indicated in the legend and explained in the text.}
    \label{fig:TaunVud}
\end{figure}

%$g_A^{ud}=1\rightarrow g_A^{pn}\approx1.27$. The FLAG 2021 average of lattice QCD calculations with $2+1+1$ flavors reads $g_A=-1.246(28)$~\cite{Aoki:2021kgd}, with the most precise single calculation predicts $g_A=-1.2642(93)$~\cite{Walker-Loud:2019cif}. 
We refer the reader to a recent review of the current experimental status of the neutron $\beta$ decay \cite{Dubbers:2021wqv}.
We summarize the existing extractions of $V_{ud}$, neutron lifetime and $\lambda$ in Fig.~\ref{fig:TaunVud}.
PDG averages~\cite{ParticleDataGroup:2022pth} are shown for $\tau_n$ (vertical green band, $\tau_n^\mathrm{av}=878.4(5)$\,s, averaged over measurements in Refs.\cite{UCNt:2021pcg,Ezhov:2014tna,Pattie:2017vsj,Serebrov:2017bzo,Arzumanov:2015tea,Steyerl:2012zz,Pichlmaier:2010zz,Serebrov:2004zf}) and $\lambda$ (wide light-blue diagonal band, $\lambda^\mathrm{av}=-1.2754(13)$, averaged over measurements in Refs.\cite{Hassan:2020hrj,Beck:2019xye,Markisch:2018ndu,UCNA:2017obv,Mund:2012fq,Schumann:2007hz,Mostovoi:2001ye,Liaud:1997vu,Erozolimsky:1997wi,Bopp:1986rt}) as indicated on the plot.
Alongside, the single most precise measurements of $\tau_n$ with the bottle method \cite{UCNt:2021pcg} (orange vertical band, $\tau_n^{\mathrm{UCN}\tau}=877.75(34)$\,s), that with the beam method \cite{Yue:2013qrc}
(yellow vertical band, $\tau_n^{\mathrm{Beam}}=887.7(2.2)$\,s), and 
$\lambda$ \cite{Markisch:2018ndu}(dark blue diagonal band, $\lambda^\mathrm{PERKEOIII}=-1.27642(56)$) are shown. The beam-bottle discrepancy in $\tau_n$ is elucidated by the distance between the yellow, and orange and green vertical bands. The Cabibbo angle anomaly (CAA) corresponds to the non-overlaping of the horizontal cyan ($0^+-0^+$ nuclear decays) and grey (unitarity constraint plus $V_{us}$ from kaon decays~\cite{ParticleDataGroup:2022pth}) bands. It is seen as well that, given the measurements of $\lambda$, the lifetime from bottle experiments is compatible with $V_{ud}$ from superallowed nuclear decays and with unitarity, 
while that from beam experiments suggests a significant discrepancy with both, as pointed out in \cite{Czarnecki:2018okw}.
Anticipating the upcoming improvement in the experimental uncertainties (and assuming the bottle lifetime value), we observe that neutron decay 
will start resolving CAA in the near future. 

Apart from the purely experimental precision which has tremendously improved over the recent years, the accuracy of $V_{ud}$ as obtained from neutron decay is limited by the theory uncertainties, stemming from the strong interaction governed by Quantum Chromodynamics (QCD) in its non-perturbative regime. These enter the proportionality coefficient in Eq.\eqref{eq:Vudn} and amount to $\approx10^{-4}$ uncertainty to $V_{ud}$, following a recent reevaluation of the SM radiative corrections (RC) with improved precision. A comprehensive overview of these developments is the main focus of this work. 

This review is organized as follows. We start with a discussion of the observables in neutron decay in Section \ref{sec:diffrate}. We lay out the structure of the SM RC in Section \ref{sec:inner} and concentrate on the $\gamma W$-box in Section \ref{sec:gammaW}. Its evaluations within the effective field theory and lattice QCD frameworks are addressed in Sections \ref{sec:EFT} and \ref{sec:LQCD}, respectively. The consequences for the new physics searches are reviewed in Section \ref{sec:BSM}, upon which we conclude with Section \ref{sec:conclusions}.

%If upon including all SM effects the inconsistencies are detected in such a global analysis, they can be interpreted in terms of physics beyond the SM (BSM), as these contributions remove the SM-based rigid correlations among various observables. Heavy BSM contributions will affect both beta decay data and collider observables alike. A review of BSM constraints on the Wilson coefficients of the dimension-6 effective 4-fermion operators is given in Section [xxx]. 

\section{Differential decay rate}\label{sec:diffrate}

%In this section we briefly outline the general structure of the $n\rightarrow pe\nu_e$ differential decay rate. The tree-level decay is triggered by the following charged weak (CW) current:
%\begin{equation}
%J_W^\mu=\Bar{u}\gamma^\mu(1-\gamma_5)d~,
%\end{equation}
%where we have scaled out the factor $V_{ud}$ for convenience. 
We are interested in the neutron decay process: $n\rightarrow p+e+\bar{\nu}_e$, where the decay kinematics can be found in Appendix~\ref{sec:kinematics}. At tree level, it probes the single-nucleon charge-current matrix element which, beyond the leading terms in Eq. \eqref{eq:Lpn_leading}, reads
\begin{align}
F^\mu\equiv\langle p(p_p,s_p)|J_W^\mu(0)|n(p_n,s_n)\rangle &=\bar{u}_{s_p}(p_p)\bigg[F_1^W\gamma^\mu+\frac{i}{2M}F_2^W\sigma^{\mu\nu}(p_p-p_n)_\nu\nonumber\\
&+G_A\gamma^\mu\gamma_5-\frac{G_P}{2M}\gamma_5(p_p-p_n)^\mu\bigg]u_{s_n}(p_n)\,.\label{eq:CCME}
\end{align}
$F_1^W$, $F_2^W$, $G_A$ and $G_P$ are the vector, weak magnetism, axial and pseudoscalar form factors which are functions of $t=(p_p-p_n)^2$, and $M\equiv (M_n+M_p)/2$ is the average nucleon mass. 

Given the small nucleon mass difference, recoil effects $\sim 10^{-3}$ are small (but important for precision~\cite{Holstein:1974zf}). 
In the non-recoil limit, only the vector and axial form factors are relevant. In particular, we define $\mathring{g}_V\equiv F_1^W(0)$ and $\mathring{g}_A\equiv G_A(0)$ as the neutron vector and axial coupling constants (we follow the experimentalists' convention and take $\mathring{g}_A<0$); here we use the upper circle to denote the ``unrenormalized'' couplings, i.e. those coming from purely strong dynamics. Conserved vector current (CVC)~\cite{Feynman:1958ty} entails  $\mathring{g}_V=1$ in the isospin limit. The deviation from unity requires strong isospin-symmetry breaking (ISB), and scales quadratically with the small ISB parameter according to the Behrens-Sirlin-Ademollo-Gatto (BSAG) theorem~\cite{Behrends:1960nf,Ademollo:1964sr}. We refer the reader to Refs.\cite{Donoghue:1990ti,Guichon:2011gc,Crawford:2022yhi} for numerical estimates, and to Ref.\cite{Seng:2023jby} for a proposed strategy of computing this deviation in lattice QCD. In turn, the axial current is not conserved, and $\mathring{g}_A$ significantly deviates from $-1$. Recent lattice QCD calculations~\cite{Gupta:2018qil,Chang:2018uxx,Walker-Loud:2019cif,Liang:2018pis,Harris:2019bih,Lin:2018obj} led to percent-level determinations~\cite{FlavourLatticeAveragingGroupFLAG:2021npn}: $\mathring{g}_A=-1.246(28)$ for $N_f=2+1+1$ and $-1.248(23)$ for $N_f=2+1$.
The couplings multiplying the recoil corrections are numerically large. The weak magnetism in the exact isospin limit is given by the isovector nucleon magnetic moment, $F_2^W(0)=\mu^p-\mu^n=4.70589007(45)$. 
The pseudoscalar coupling is enhanced, $G_P(0)=-(M/\bar m_q)\mathring{g}_A=349(9)$, with $\bar m_q$ the average light quark mass \cite{Gonzalez-Alonso:2013ura}, but the quadratic dependence on the small recoil suppresses it beyond the current precision level.

The differential decay rate of a polarized neutron to unpolarized final states takes the following form~\cite{Jackson:1957zz,Sirlin:1967zza,Garcia:1978bq}:
\begin{align}
    \left(\frac{d\Gamma}{dE_e d\Omega_ed\Omega_\nu}\right)_0&=\frac{(G_FV_{ud})^2}{(2\pi)^5}F(E_e)|\vec{p}_e|E_e(E_0-E_e)^2(1+3\lambda^2)g_V^2\left(1+\frac{\alpha_\text{em}}{2\pi}\delta^{(1)}(E_e)\right)\nonumber\\
    &\times\left( 1+
    (a+\delta a_r)\left(1+\frac{\alpha_\text{em}}{2\pi}\delta^{(2)}(E_e)\right)\frac{\vec{p}_e\cdot\vec{p}_\nu}{E_eE_\nu}+b\frac{m_e}{E_e}\right.\label{eq:drate}\\    &\left.+\hat{\sigma}\cdot\left[
 (A+\delta A_r)
 \left(1+\frac{\alpha_\text{em}}{2\pi}\delta^{(2)}(E_e)\right)\frac{\vec{p}_e}{E_e}+
 (B+\delta B_r)
 \frac{\vec{p}_\nu}{E_\nu}
    +D\frac{\vec p_e\times\vec p_\nu}{E_eE_\nu}\right]\right)~.\nonumber
\end{align}
Above,
$\hat{\sigma}$ is the unit neutron polarization vector, and $E_0=(M_n^2-M_p^2+m_e^2)/(2M_n)$ is the electron end-point energy. %; extra correlation structures which are suppressed by recoil corrections are not displayed above. 
The quantities $a,b,A,B,D$ are referred to as correlation coefficients, and $\{\delta a_r,\delta A_r,\delta B_r\}$ are the known recoil corrections ($\sim 1/M$) thereto, generated by the nucleon magnetic moment~\cite{Cirigliano:2022hob}.
If neglecting RC and recoil corrections, the non-vanishing coefficients read:
\begin{equation}
 a_0=\frac{1-\lambda_0^2}{1+3\lambda_0^2}~,~A_0=\frac{-2\lambda_0-2\lambda_0^2}{1+3\lambda_0^2}~,~B_0=\frac{-2\lambda_0+2\lambda_0^2}{1+3\lambda_0^2}~,\label{eq:barecoef}
\end{equation}
where $\lambda_0=\mathring{g}_A/\mathring{g}_V$ is the ratio of the bare axial and vector coupling constants. The expressions above suggest that one could measure the axial coupling $\mathring{g}_A$ either through recoil effects ($a_0$) or the $\hat{\sigma}$-dependence ($A_0,B_0$) in the differential decay rate. 

To achieve $10^{-3}$ precision it is necessary to include electromagnetic and recoil effects. 
The electromagnetic corrections to neutron decay are of two kinds. The first are $E_e$-dependent terms that distort the beta spectrum. The largest of such kind is the Fermi function $F(E_e)$ which describes the Coulomb interaction between the final-state proton and electron~\cite{Fermi:1934hr}. The remaining $E_e$-dependent terms not included in the Fermi function are  collectively coined as ``outer RC'' and appear as the corrections $\delta^{(1,2)}(E_e)$ in Eq.\eqref{eq:drate}.\footnote{Notice: $\left(1+\frac{\alpha_\text{em}}{2\pi}\delta^{(1)}_\alpha(E_e)\right)$ in Ref.\cite{Cirigliano:2022hob} corresponds to  $g_V^2\left(1+\frac{\alpha_\text{em}}{2\pi}\delta^{(1)}(E_e)\right)$ in this work.} 
The relevant expressions can be found in, e.g. Ref.\cite{Seng:2021syx}. The further corrections that are $E_e$-independent are known as ``inner RC'', and they modify the coupling constants in the tree-level charged weak matrix element as
\begin{equation}
\mathring{g}_V\rightarrow\mathring{g}_V+\delta g_V\equiv g_V ~,~\mathring{g}_A\rightarrow\mathring{g}_A+\delta g_A\equiv g_A~.
\end{equation}
They renormalize the total decay rate by an overall factor $g_V^2$, and the correlation coefficients $a,A,B$ are modified by replacing $\lambda_0$ in Eq.\eqref{eq:barecoef} with the renormalized ratio $\lambda\equiv g_A/g_V$. 

The coefficient $b$ is often referred to as Fierz term, and requires an effective scalar interaction. In the SM it is double-suppressed, $b\sim\alpha m_e/M_p\sim10^{-6}$ and hence is a promising avenue to look for non-standard scalar and tensor currents~\cite{Erler:2004cx,Bhattacharya:2011qm} (see Refs.\cite{Hickerson:2017fzz,Saul:2019qnp} for limits from neutron decay). The coefficient $D$ is a time reversal-odd effect~\cite{Jackson:1957zz} that can be generated in SM by the final-state interaction effects, such as the Coulomb phase~\cite{Jackson:1957auh}. While beyond-the-Standard-Model (BSM) $CP$-violating contributions to $D$ are severely constrained by the neutron electric dipole moment (EDM) experiments, $CP$-conserving BSM effects in interference with the SM Coulomb phase may generate $D\sim10^{-5}$ testable in the upcoming experiments~\cite{Callan:1967zz,Falkowski:2022ynb}.

Upon integrating over the phase space, the full decay rate $\Gamma=1/\tau_n$ can be used to determine $|V_{ud}|$ if we rewrite Eq.\eqref{eq:Vudn} in a more detailed manner~\cite{Zyla:2020zbs}, 
\begin{equation}
|V_{ud}|_n^2=\frac{5024.7~\text{s}}{\tau_n(1+3\lambda^2)(1+\Delta_R^V)}~.\label{eq:Vudmaster}
\end{equation}
The numerator at the right hand side contains the effects of the Fermi constant $G_F$ measured from muon decay and the statistical rate function $f$~\cite{Wilkinson:1982hu,Hayen:2017pwg}, as well as the outer RC. Beyond the neutron lifetime $\tau_n$ and the renormalized ratio $\lambda$ (both experimentally measured), 
the denominator contains the theoretical quantity $\Delta_R^V$ representing the inner RC to the vector coupling. This quantity will be discussed in detail below. 

To summarize, neutron decay serves a valuable tool to test internal consistency of SM. In the electroweak sector, precise measurements of the neutron lifetime and correlation coefficients (i.e., the renormalized axial coupling $g_A$) provide an accurate way of extracting $V_{ud}$ which can be compared with that obtained from nuclear and pion decays. Combined furthermore with other CKM matrix elements extracted from heavier quark flavors decays, it allows for highly accurate tests of unitarity of the CKM matrix, a cornerstone of SM. 
In the strong interaction sector, the bare axial coupling constant $\mathring{g}_A$ obtained from the measured correlation coefficients can be confronted to the ever more precise lattice QCD calculations. 
For both tests,
precise knowledge of RC which should be removed from the measured quantities for a meaningful comparison, is mandatory.

\section{Inner radiative corrections\label{sec:inner}}
\begin{figure}[t]
\begin{center}
\includegraphics[width=0.7\columnwidth]{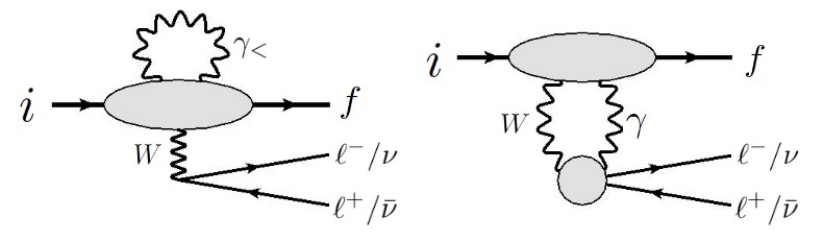}
\caption{One-loop Feynman diagrams of electroweak RC that probe non-perturbative strong interactions.}
\label{fig:NPRC}
\end{center}
\end{figure}

The first systematic analysis of RC fully compatible with the SM electroweak theory was established in 1978 by Sirlin~\cite{Sirlin:1977sv}. It is not the purpose of this review to repeat Sirlin's derivation from scratch, and interested readers are referred to more comprehensive reviews~\cite{Sirlin:2012mh,Seng:2021syx}. Here we jump directly to the most important conclusion: Sirlin showed that, among all the $\mathcal{O}(\alpha_\text{em})$ RC to a generic semileptonic beta decay $\phi_i\rightarrow\phi_f$ (which could be either $\beta^+$ or $\beta^-$), the only two diagrams that depend on non-perturbative strong interactions are those in Fig.\ref{fig:NPRC}. The first diagram represents the RC to the charged current matrix element (e.g., Eq.\eqref{eq:CCME} for the case of single nucleon), where $\gamma_<$ denoting the photon propagator with a Pauli-Villars regulator $M_W^2/(M_W^2-q^2)$, and the second diagram is the  $\gamma W$-box diagram. 

Using Ward identity, Sirlin further showed that the first diagram splits into two pieces:
\begin{equation}
    \delta F^\mu=\delta F_\text{2pt}^\mu+\delta F_\text{3pt}^\mu~,\label{eq:Fmusplit}
\end{equation}
which we name as the ``two-point function'' and ``three-point function'' respectively. The two-point function takes the following form for neutron:
\begin{equation}
    \delta F_{\text{2pt}}^\lambda=\frac{e^2}{2}\int \frac{d^4q}{(2\pi)^4}\frac{\partial}{\partial q_\lambda}\left(\frac{M_W^2}{M_W^2-q^2}\frac{1}{q^2-m_\gamma^2}\right)T^\mu_{\gamma W\mu}~,\label{eq:2pt}
\end{equation}
where\footnote{The normalization of $T_{\gamma W}^{\mu\nu}$ here is the same as Ref.\cite{Seng:2021syx}, but is two times as large as Refs.\cite{Seng:2018qru,Seng:2018yzq}.}
\begin{equation}
    T_{\gamma W}^{\mu\nu}\equiv \int d^4x e^{iq\cdot x}\langle p|T[J_\text{em}^\mu(x)J_W^\nu(0)]|n\rangle
\end{equation}
is a ``generalized Compton tensor'' consists of the time-order product of the electromagnetic and charged weak current. The definition of the ``three-point function'' is more complicated and consists of two terms:
\begin{equation}
    \delta F_\text{3pt}^\lambda=-\lim_{\delta\bar{p}\rightarrow \delta p}i\delta\bar{p}_\nu\frac{\partial}{\partial\delta\bar{p}_\lambda}\left[\bar{T}^\nu-B^\nu\right]+\lim_{\delta\bar{p}\rightarrow \delta p}i\frac{\partial}{\partial\delta\bar{p}_\lambda}\left[\mathbb{D}-\delta\bar{p}\cdot B\right]~,
\label{eq:3pt}
\end{equation}
where $\delta p\equiv p_n-p_p$, and 
\begin{eqnarray}
\bar{T}^\mu&=&\frac{e^2}{2}\int\frac{d^4q}{(2\pi)^4}\frac{M_W^2}{M_W^2-q^2}\frac{1}{q^2-m_\gamma^2}\int d^4x e^{i\delta\bar{p}\cdot x} d^4y e^{iq\cdot y}\langle p|T\{J_W^\mu(x)J_\text{em}^\nu(y)J_\nu^\text{em}(0)\}|n\rangle\nonumber\\
\mathbb{D}&=&\frac{ie^2}{2}\int\frac{d^4q}{(2\pi)^4}\frac{M_W^2}{M_W^2-q^2}\frac{1}{q^2-m_\gamma^2}\int d^4x e^{i\delta\bar{p}\cdot x} d^4y e^{iq\cdot y}\langle p|T\{\partial\cdot J_W(x)J_\text{em}^\nu(y)J_\nu^\text{em}(0)\}|n\rangle\nonumber\\
B^\mu&=&-\bar{u}_p\left[\frac{i\delta M_p}{\slashed{p}_n-\delta\bar{\slashed{p}}-M_p}\mathfrak{T}^\mu+\mathfrak{T}^\mu\frac{i\delta M_n}{\slashed{p}_p+\delta\bar{\slashed{p}}-M_n}\right]u_n~,
\end{eqnarray}
with $\delta M_{p,n}$ the nucleon mass shift due to electromagnetic corrections, and $\mathfrak{T}^\mu$ the nucleon vertex function. The $B$-terms in Eq.\eqref{eq:3pt} removes the poles in $\bar{T}^\nu$ and $\mathbb{D}$ at $\delta\bar{p}\rightarrow\delta p$.    

An important observation is that $\delta F_\text{2pt}^\mu$ partially cancels the part of the $\gamma W$-box diagram which stems from the symmetric $T_{\gamma W}^{\mu\nu}$, leaving a residual piece that depends only on IR-physics and can be evaluated analytically. This residual piece gives rise to the outer correction which was known well before the birth of SM~\cite{Kinoshita:1957zz,Kinoshita:1958ru,Sirlin:1967zza}. The uncancelled piece of the box diagram, coming from the antisymmetric part of $T_{\gamma W}^{\mu\nu}$, contributes to the inner correction to $g_{V,A}$ as~\cite{Gorchtein:2021fce}  
\begin{eqnarray}
\frac{\delta g_V^{\gamma W}}{\mathring{g}_V}\equiv\Box_{\gamma W}^V&=&\frac{e^2}{2M\mathring{g}_V}\int\frac{d^4q}{(2\pi)^4}\frac{M_W^2}{M_W^2+Q^2}\frac{1}{(Q^2)^2}\frac{\nu^2+Q^2}{\nu}T_3\nonumber\\
\frac{\delta g_A^{\gamma W}}{\mathring{g}_A}\equiv\Box_{\gamma W}^A&=&\frac{e^2}{M\mathring{g}_A}\int\frac{d^4q}{(2\pi)^4}\frac{M_W^2}{M_W^2+Q^2}\frac{1}{(Q^2)^2}\left\{\frac{\nu^2-2Q^2}{3\nu}S_1-\frac{Q^2}{\nu}S_2\right\}~,\label{eq:box}
\end{eqnarray}
where the invariant amplitudes $T_3$, $S_{1,2}$ are defined through $T_{\gamma W}^{\mu\nu}$ in the forward limit (i.e. $p_n=p_p=p$): 
\begin{equation}
    T_{\gamma W}^{\mu\nu}=-\frac{i\epsilon^{\mu\nu\alpha\beta}q_\alpha p_\beta}{2p\cdot q}T_3+\frac{i\epsilon^{\mu\nu\alpha\beta}q_\alpha}{p\cdot q}\left[S_\beta S_1+\left(S_\beta-\frac{S\cdot q}{p\cdot q}p_\beta\right)S_2\right]+...,
\end{equation}
with $Q^2=-q^2$, $\nu=p\cdot q/M$, and $S^\mu$ is the nucleon's spin vector normalized as $S^2=-M^2$. 

Finally, let us address the properties of the three-point functions. The first term in Eq.\eqref{eq:3pt} vanishes for $\delta p\rightarrow 0$ and can safely be neglected for neutron decay. The second term is more complicated: it vanishes if $\partial\cdot J_W=0$, i.e. if the charged weak current is conserved. For the vector component, this is just the CVC hypothesis which is an excellent approximation given that $(\partial\cdot J_W)_V\sim (m_d-m_u)\sim \delta p$. One might naively think that the same works for the axial current given the partially-conserved axial current (PCAC) relation $(\partial\cdot J_W)_A\sim m_\pi^2 $; but since in reality one has $\delta p\ll m_\pi$, a valid approximation is instead to drop terms that scale as $\delta p/m_\pi$. When such approximation is made, the axial current becomes explicitly non-conserved, and its contribution to $\delta F_\text{3pt}^\mu$ is non-zero. Therefore, $\delta F_\text{3pt}^\mu$ may contribute to the outer correction $\delta^{(2)}$ and the inner correction to $g_A$, but not to $g_V$. 

To summarize, the renormalized vector and axial coupling can be expressed as:
\begin{eqnarray}
    g_V^2&=&\mathring{g}_V^2\left\{1+\Delta_R^V\right\}=\mathring{g}_V^2\left\{1+\Delta_R^U+2\Box_{\gamma W}^V\right\}\nonumber\\
    g_A^2&=&\mathring{g}_A^2\left\{1+\Delta_R^A\right\}=\mathring{g}_A^2\left\{1+\Delta_R^U+2\Box_{\gamma W}^A+\Delta_{R,\text{3pt}}^{A}\right\}~,
\end{eqnarray}
which defines the quantity $\Delta_R^V$ that appears in Eq.\eqref{eq:Vudmaster}. In the above, 
\begin{equation}
\Delta_R^U=\frac{\alpha_{em}}{2\pi}\left[3\ln\frac{M_z}{M_p}+\ln\frac{M_Z}{M_W}+\tilde{a}_g\right]+\delta^\text{QED}_\text{HO}= 0.01709(10)\label{eq:universal}
\end{equation}
is a universal piece that consists of the analytically-calculable ``weak'' RC, the pQCD correction (not coming from $\Box_{\gamma W}$) $\tilde{a}_g\approx -0.083$~\cite{Seng:2021syx}, the resummation of leading QED logarithms (from all diagrams including $\Box_{\gamma W}$) and the most important $\mathcal{O}(\alpha_\text{em}^2)$ corrections $\delta^\text{QED}_\text{HO}= 0.00109(10)$~\cite{Czarnecki:2004cw}, where we include a $\pm 1\times 10^{-4}$ uncertainty following Ref.\cite{Marciano:2005ec}. With this, the renormalized axial-vector ratio $\lambda$ is related to the bare ratio as:
\begin{equation}
    \lambda^2=\lambda_0^2\left\{1+2\Box_{\gamma W}^A-2\Box_{\gamma W}^V+\Delta_{R,\text{3pt}}^A\right\}~.
\end{equation}
It is worthwhile noting that for a long time the three-point function $\delta F_{\text{3pt}}^\mu$ was mistakenly regarded to be small, which caused the incorrect neglect of $\Delta_{R,\text{3pt}}^A$ in the literature, including our own works~\cite{Gorchtein:2021fce,Seng:2021syx}. 
A recent analysis in the effective field theory (EFT) framework \cite{Cirigliano:2022hob} suggests that contributions to $\Delta_{R,\text{3pt}}^A$ as large as 1-2\% cannot be excluded. We review this exciting new development in detail in Section \ref{sec:EFT}.

\section{$\gamma W$-box diagram in dispersive representation\label{sec:gammaW}}

In this section we discuss the single-nucleon $\gamma W$-box diagram correction to $g_V$ and $g_A$. A long-standing problem in the area of precision physics, it has recently been reevaluated in a new framework which allowed to reduce its uncertainty. 
Historically, one evaluated the box diagram integrals $\Box_{\gamma W}^{V,A}$ directly using Eq.\eqref{eq:box}. Since the invariant amplitudes $T_3$, $g_{1,2}$ are not directly measurable in the experiment, a substantial amount of theory modeling of the hadron dynamics at the non-perturbative scale $Q^2\sim 1$~GeV$^2$ is unavoidable. With this, controlling the overall uncertainty at the level of $10^{-4}$ is challenging. 

The use of dispersion relation (DR) provides a satisfactory solution to the problem above~\cite{Seng:2018qru,Seng:2018yzq,Shiells:2020fqp,Gorchtein:2021fce}. In this formalism, one re-expresses the box diagram integrals in terms of single-nucleon
 structure functions:
\begin{eqnarray}
\Box_{\gamma W}^V&=&\frac{\alpha_\text{em}}{\pi\mathring{g}_V}\int_0^\infty\frac{dQ^2}{Q^2}\frac{M_W^2}{M_W^2+Q^2}\int_0^1dx_B\frac{1+2r}{(1+r)^2}F_3^{(0)}\nonumber\\
\Box_{\gamma W}^A&=&-\frac{2\alpha_\text{em}}{\pi\mathring{g}_A}\int_0^\infty\frac{dQ^2}{Q^2}\frac{M_W^2}{M_W^2+Q^2}\int_0^1 \frac{dx_B}{(1+r)^2}\left\{\frac{5+4r}{3}g_1^{(0)}-\frac{4M^2x_B^2}{Q^2}g_2^{(0)}\right\}~,\nonumber\\
\end{eqnarray}
where $r=\sqrt{1+4M^2x_B^2/Q^2}$, with $x_B=Q^2/(2p\cdot q)$ the standard Bjorken variable. The structure functions are defined through:
\begin{eqnarray}
    W_{\gamma W}^{\mu\nu}&=&\frac{1}{4\pi}\sum_X(2\pi)^4\delta^{(4)}(p+q-p_X)\langle p|J_\text{em}^\mu(0)|X\rangle \langle X|J_W^\nu(0)|n\rangle\nonumber\\
    &=&-\frac{i\epsilon^{\mu\nu\alpha\beta}q_\alpha p_\beta}{2p\cdot q}F_3+\frac{i\epsilon^{\mu\nu\alpha\beta}q_\alpha}{p\cdot q}\left[S_\beta g_1+\left(S_\beta-\frac{S\cdot q}{p\cdot q}p_\beta\right)g_2\right]+...
\end{eqnarray}
with $S$ denoting the nucleon spin 4-vector, 
and the superscript $(0)$ indicating the contribution from the isosinglet component of the electromagnetic current.

\begin{figure}[t]
\begin{center}
\includegraphics[width=0.6\columnwidth]{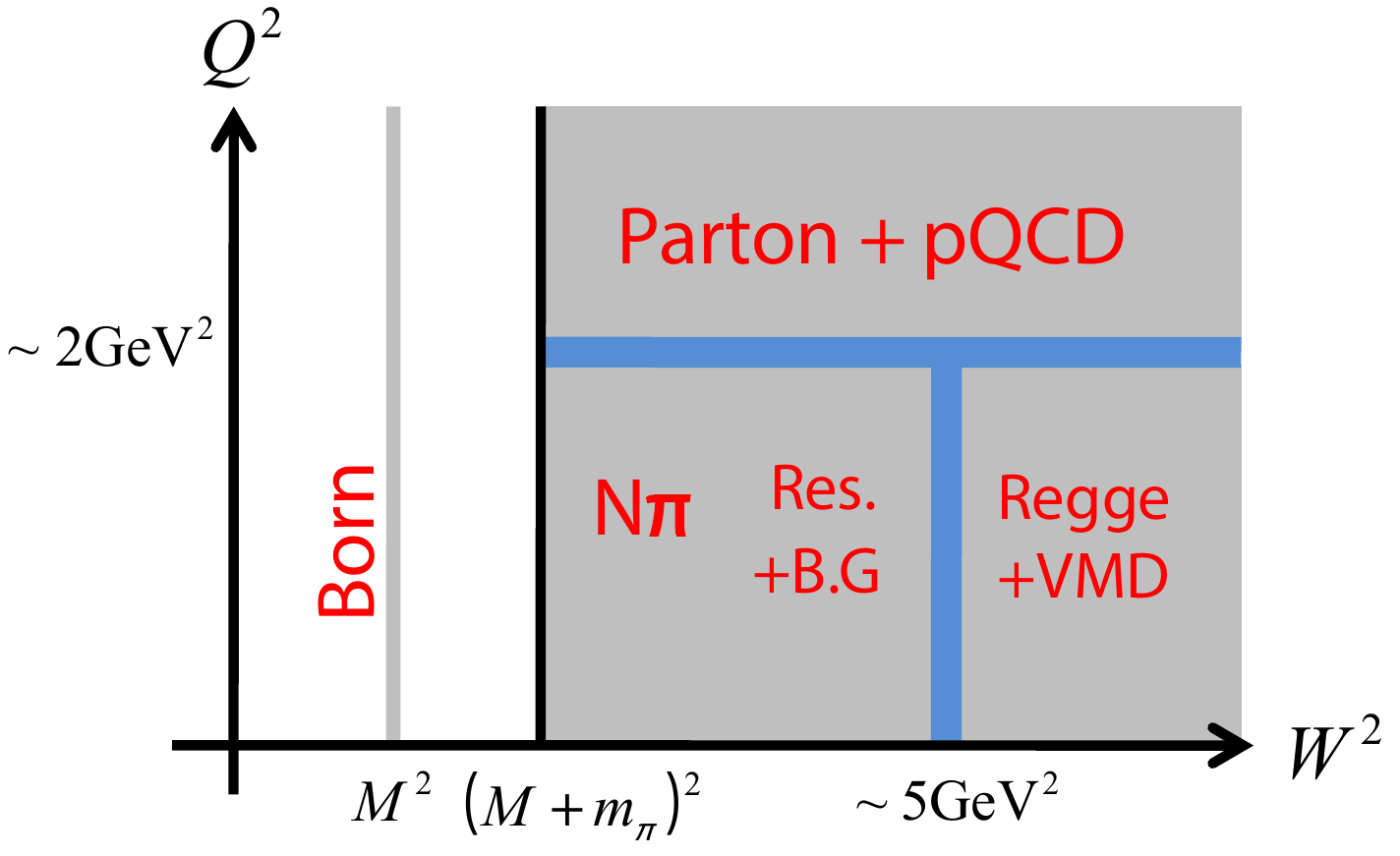}
\caption{Main contributors to the single-nucleon structure functions at different regions in the $W^2$--$Q^2$ plane. Figure adopted from Ref.\cite{Seng:2018yzq}.}
\label{fig:W-Q2diag}
\end{center}
\end{figure}

Fig.\ref{fig:W-Q2diag} depicts the main contributions to the nucleon structure functions across different kinematic regions in $Q^2$ and $W^2\equiv (p+q)^2$. Moving from the left to the right at a constant $Q^2$, the first contribution one encounters is the elastic (Born) contribution. It appears as a simple pole (a $\delta$-function in the structure functions) at $W^2=M^2$ at all values of $Q^2$ and is separated from inelastic contributions by a finite energy gap. 
Starting at $W^2\geq W_\pi^2\equiv (M+m_\pi)^2$
(represented by the solid black vertical line)
multi-hadron states become energetically allowed, with the lightest such state being a nucleon and a pion. As the energy variable $W$ further increases, nucleon and $\Delta$ resonances can be excited, in addition to the non-resonant continuum. At even higher energies high hadron multiplicity contributions gradually take over, the transition at $W^2\sim5$ GeV$^2$ represented by a broad blue vertical band. 
These multi-hadron contributions are well described in the language of the $t$-channel Regge exchanges. 

As moving from the bottom to the top of Fig.\ref{fig:W-Q2diag}, 
these contributions start to evolve. The low-energy contributions follow elastic or transition form factors $F(Q^2)$ which generically suppress the respective contributions to the structure functions as $F^2(Q^2)\sim(1+\frac{Q^2}{\Lambda^2})^{-4}$ at high $Q^2$, with $\Lambda$ a characteristic hadronic scale of the order of 1 GeV. A milder suppression is observed at high energies where the vector dominance model (VDM) picture predicts the behavior of the structure functions to roughly follow $F_\mathrm{VDM}^2(Q^2)\sim(1+\frac{Q^2}{\Lambda^2})^{-2}$. This behavior is in stark contrast with Bjorken scaling which experimental data exhibit at high $Q^2$ where $F_3(x,Q^2)\approx F_3(x)$, as predicted by the parton model. Perturbative QCD (pQCD) corrections reintroduce the $Q^2$ dependence to the structure functions, becoming increasingly important as one moves down in $Q^2$. The two patterns, the non-perturbative hadronic and the perturbative QCD ones, join smoothly around $Q^2\approx2$ GeV$^2$.

The boundaries between the various mechanisms and patterns, described above and indicated in Fig.\ref{fig:W-Q2diag} by the blue bands, are very approximate. How exactly the transition between various regions, in particular from the perturbative to the nonperturbative regime occurs, is a long-standing open problem in QCD. 
In the problem at hand, the lack of detailed understanding of this transition and the model-dependent description of the inclusive structure functions in the region in between, is the primary source of the overall uncertainty of $\Delta_R^V$. The DR formalism is handy at approaching this problem in a direct way. If sufficiently precise and abundant data on structure functions exist, the integral can be evaluated model-independently, with the uncertainty driven almost entirely by the experiment. This is the case for $\Box_{\gamma W}^A$, whereas for $\Box_{\gamma W}^V$ no direct experimental input is available. Hence, for the latter an additional uncertainty stemming from relating  the low-$Q^2$ inelastic contribution to data or other sources of input appears.

Historically,  $\Box_{\gamma W}^V$ was first to receive attention in the context of extracting $V_{ud}$ from superallowed decays. $\Box_{\gamma W}^A$ was of less interest because it affects the Gamow-Teller (GT) strength that does not offer a clean access to $V_{ud}$ because of effects of strong interaction. Here, we will proceed in a reverse chronological order, by first discussing the determination of $\Box_{\gamma W}^A$ which is largely model-independent, and later move on to $\Box_{\gamma W}^V$ which requires more modeling.

\subsection{Model-independent determination of $\Box_{\gamma W}^A$}

The procedure for a data-driven analysis of $\Box_{\gamma W}^A$ was thoroughly explained in Ref.\cite{Gorchtein:2021fce} and here we briefly summarize the most important results. The starting point is the following isospin relation:
\begin{equation}
    g_{1,2}^{(0)}=\frac{1}{2}\left\{g_{1,2}^p-g_{1,2}^n\right\}
\end{equation}
which relates the charged weak spin structure functions to the isovector combination of the corresponding electromagnetic structure functions. Since the latter can be obtained from experiments, $g_{1,2}^{(0)}$ can be determined model-independently. One may write:
\begin{equation}
    \Box_{\gamma W}^A=\frac{\alpha_\text{em}}{2\pi}[d_B+d_1+d_2]~,
\end{equation}
where the terms at the RHS represent the elastic (Born) contribution, the inelastic contribution from $g_1^{(0)}$ and that from $g_2^{(0)}$, respectively. The Born contribution to the spin-dependent structure functions read:
\begin{equation}
    g_{1,B}^{(0)}=\frac{F_1^VG_M^S+F_1^S G_M^V}{8}\delta(1-x_B)~,~g_{2,B}^{(0)}=-\tau\frac{F_2^VG_M^S+F_2^SG_M^V}{8}\delta(1-x_B)~,
\end{equation}
where $G_E=F_1-\tau F_2$, $G_M=F_1+F_2$ are the Sachs form factors, with $\tau=Q^2/(4M^2)$. The isovector (V) and isoscalar (S) nucleon form factors are related to the proton and neutron electromagnetic form factors as: $A^V=A^p-A^n$, $A^S=A^p+A^n$. The latter have been studied extensively~\cite{Drechsel:2002ar,Lorenz:2012tm,Lorenz:2014yda,Ye:2017gyb,Lin:2021umk,Lin:2021umz,Lin:2021xrc}, leading to a determination of the Born contribution with a 0.8\% precision, $d_B=1.22(1)$.  

\begin{figure}[t]
\begin{center}
\includegraphics[width=0.4\columnwidth]{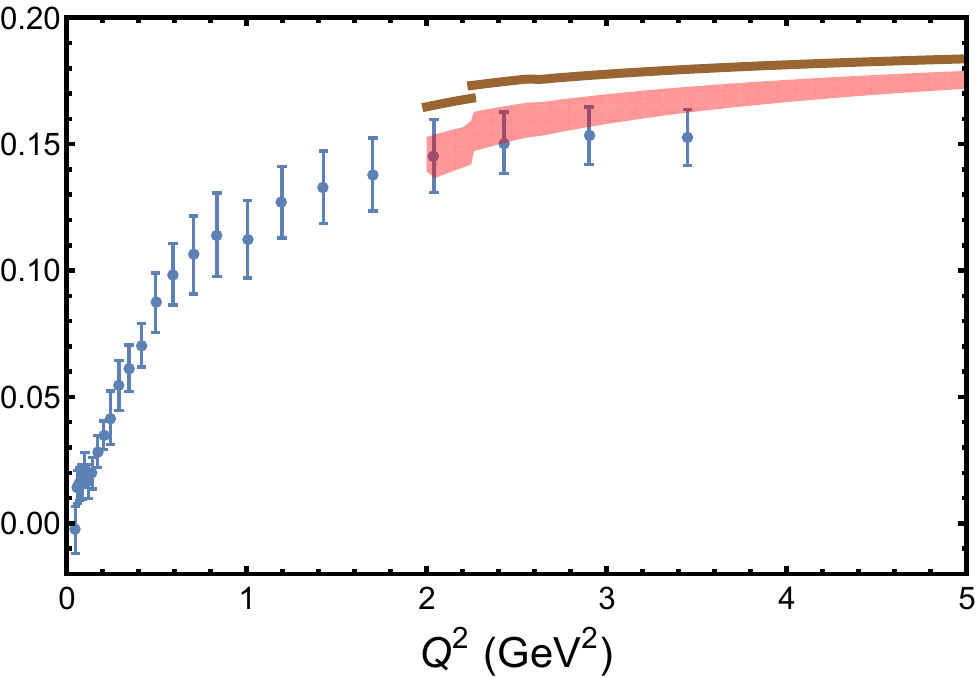}
\caption{$\bar{\Gamma}_1^{p-n}$ reconstructed from data at small $Q^2$ versus theory prediction at large $Q^2$, with (red band) and without (brown curve) higher-twist corrections. Figure adopted from Ref.\cite{Gorchtein:2021fce}.}
\label{fig:Gammabar1}
\end{center}
\end{figure}

Inelastic contribution at $W^2\geq W_\pi^2$ should be evaluated separately for $Q^2>Q_0^2$ and $Q_2<Q_0^2$, where $Q_0^2$ is a momentum scale above which perturbative description applies. A common choice is $Q_0^2=2~$GeV$^2$ which will later be justified by data. To proceed, one needs to first understand what data we have; we start from the dominant contribution which comes from $g_1^{(0)}$. The EG1b experiment by the CLAS collaboration at Jefferson Lab provided measurements of the first few Mellin moments of the structure function $g_1$ for proton and neutron at $0.05~\text{GeV}^2<Q^2<5~\text{GeV}^2$~\cite{CLAS:2015otq,CLAS:2017qga}:
\begin{equation}
    \Gamma_i^N(Q^2)\equiv \int_0^{x_\pi}dx_B x_B^{i-1}g_1^N(x_B,Q^2)~,~N=p,n
\end{equation}
where $x_\pi= Q^2/(W_\pi^2-M^2+Q^2)$. 
On the other hand, the box diagram contribution depends on the following integral:
\begin{equation}
    \bar{\Gamma}_1^{p-n}(Q^2)\equiv \int_0^{x_\pi}dx_Bf(x_B,Q^2)\left\{g_1^p(x_B,Q^2)-g_1^n(x_B,Q^2)\right\}
\end{equation}
where $f(x_B,Q^2)=4(5+4r)/(9(1+r)^2)$. When $Q^2\rightarrow\infty$, we have $f(x,Q^2)\rightarrow 1$, but at small and moderate $Q^2$ the target mass corrections are important. Fortunately, for each discrete value of $Q^2$ one could perform the following fit:
\begin{equation}
    f(x_B,Q^2)\approx a(Q^2)+b(Q^2)x_B^2+c(Q^2)x_B^4~,~0<x_B<x_\pi
\end{equation}
to determine the coefficients $a(Q^2)$, $b(Q^2)$ and $c(Q^2)$ (notice that this is NOT a simple Taylor expansion of $x_B^2M^2/Q^2$, as the latter emphasizes too much on $x_B\approx 0$ and performs poorly at $x_B\rightarrow x_\pi$). With that, one is able to precisely reconstruct $\bar{\Gamma}_1^{p-n}(Q^2)$ at $Q^2<Q_0^2$ using the data of the first few Mellin moments $\Gamma_i^{p-n}$ ($i=1,3,5$). Meanwhile, for $Q^2>Q_0^2$ we have $\bar{\Gamma}_1^{p-n}\approx \Gamma_1^{p-n}$, and the latter is a sum of the leading and higher-twist contributions:
\begin{equation}
    \Gamma_1^{p-n}(Q^2)=\frac{|\mathring{g}_A|}{6}C_\text{Bj}(Q^2)+\sum_{i=2}^\infty\frac{\mu_{2i}^{p-n}}{Q^{2i-2}}~,~Q^2>Q_0^2\label{eq:Gamma1pn}
\end{equation}
The leading-twist coefficient $C_\text{Bj}(Q^2)$ satisfies the pQCD-corrected~\cite{Baikov:2010iw,Baikov:2010je} polarized Bjorken sum rule (BjSR)~\cite{Bjorken:1966jh,Bjorken:1969mm}:
\begin{equation}
    C_\text{Bj}(Q^2)=1-\sum_{n=1}^\infty \tilde{c}_n\left(\frac{\alpha_s}{\pi}\right)^n~,
    \label{eq:BjSRpQCD}
\end{equation}
while the higher-twist coefficients $\mu_{2i}^{p-n}$ (we only include $i=2$) are generally model dependent \cite{Deur:2014vea,Kotlorz:2017wpu,Ayala:2018ulm}. 
With the treatment above, one is able to construct $\bar{\Gamma}_1^{p-n}$ for all values of $Q^2$, see Fig.\ref{fig:Gammabar1}. The good agreement between theory prediction and experimental data at $Q_0^2=2$~GeV$^2$ justifies our choice for the matching point. This leads to $d_1=2.14(4)_\text{data}(1)_\text{HT}$, a very robust result with a mere 2\% uncertainty.  

The contribution of $g_2$ is much smaller as a consequence of the Burkhardt-Cottingham sum rule that requires the first moment of $g_2$ (the sum of the elastic and inelastic contributions) to vanish identically~\cite{Burkhardt:1970ti}. The remnant can be estimated with the use of the Wandzura-Wilczek relation~\cite{Wandzura:1977qf} for the leading twist, and baryon chiral effective field theory~\cite{Alarcon:2020icz} for the higher twist. The result reads $d_2=0.05(3)$. Combining all pieces we arrive at
\begin{equation}
    \Box_{\gamma W}^A=3.96(6)\times 10^{-3}~,
\end{equation}
with the uncertainty well below $10^{-4}$. We can compare this result to that of  Hayen~\cite{Hayen:2020cxh} which appeared earlier. We observe that the value of the HT parameter 
$\mu_4=-0.047(22)$ GeV$^2$ adopted from Ref.~\cite{Kotlorz:2017wpu} and used in our work,
 is an order of magnitude larger than what Hayen used. Its effect on the central value is largely compensated by the fact that Ref.~\cite{Hayen:2020cxh} neglected the $d_2$ contribution, making our central value close to Hayen's, albeit with some tension, $\Box_{\gamma W}^A=4.11(9)\times10^{-3}$. The larger uncertainty in the estimate of Ref.~\cite{Hayen:2020cxh} stems from using a model to describe the low-$Q^2$ region of the integrand, rather than the direct experimental data, which we deem unnecessary.
 
\subsection{Dispersive determinations of $\Box_{\gamma W}^V$}

Next we turn to the vector box diagram. We first write:
\begin{equation}
    \Box_{\gamma W}^V=\frac{3\alpha_\text{em}}{4\pi}\int_0^\infty\frac{dQ^2}{Q^2}\frac{M_W^2}{M_W^2+Q^2}M_3^{(0)}(1,Q^2)~,\label{eq:BoxinNachtmann}
\end{equation}
where
\begin{equation}
    M_3^{(0)}(1,Q^2)\equiv \frac{4}{3\mathring{g}_V}\int_0^1dx_B\frac{1+2r}{(1+r)^2}F_3^{(0)}(x_B,Q^2)
\end{equation}
is the first Nachtmann moment of $F_3^{(0)}$~\cite{Nachtmann:1973mr,Nachtmann:1974aj}; at large $Q^2$ it reduces to the simpler, first Mellin moment, but at small $Q^2$ it incorporates the effect of the finite target mass.

The treatment of the vector $\gamma W$-box diagram is more complicated as it is not so straightforward to identify the pertinent experimental data for the parity-odd structure function $F_3^{(0)}$. In principle, one could make use of the following isospin relation:
\begin{equation}
    F_3^{(0)}=\frac{1}{2}\left\{F_{3,\gamma Z}^p-F_{3,\gamma Z}^n\right\}
\end{equation}
where $F_{3,\gamma Z}^N$ ($N=p,n$) is the spin-independent, parity-odd structure function. It originates from the interference between the electromagnetic and neutral weak current and is accessible, at least in principle, with inclusive parity-violating electron scattering experiments on light nuclei. Such experiments are notoriously difficult and only very limited (in precision and kinematical coverage) data on $F_{1,2,\gamma Z}^{p,n}$ at low $Q^2$ exist~\cite{QWeak:2019kdt,Wang:2014guo}, but not on $F_{3,\gamma Z}^{p,n}$.

In absence of direct constraints from data, the  decomposition into distinct physical mechanisms illustrated in Fig.\ref{fig:W-Q2diag} comes in handy. It provides a useful guidance to separate the low-$Q^2$ contribution into two classes:
\begin{itemize}
    \item ``Non-asymptotic'' pieces (Born, low-energy continuum, resonances) that are different for different channels of $F_3$ and need to be calculated case-by-case; 
    \item ``Asymptotic'' piece at high energy which is universal for different channels of $F_3$ (up to Clebsch-Gordon factors). This piece can be extracted from experimental data or other measurable structure functions. Within this latter class we also distinguish the ``subasymptotic" part which, while being largely universal, contains a significant amount of model dependence.
\end{itemize}

In the post-2018 works that study $\Box_{\gamma W}^V$~\cite{Seng:2018qru,Seng:2018yzq,Shiells:2020fqp,Seng:2020wjq,Czarnecki:2019mwq,Hayen:2020cxh}, the treatment of the non-asymptotic pieces is similar. The Born contribution is fixed by the experimental data of the nucleon axial and magnetic Sachs form factors~\cite{Ye:2017gyb,Lorenz:2012tm,Lorenz:2014yda,Lin:2021umk,Lin:2021umz,Lin:2021xrc,Bhattacharya:2011ah}. The low-energy $N\pi$ and resonance contribution can be estimated using chiral perturbation theory and existing parameterization of resonance matrix elements~\cite{Lalakulich:2006sw,Drechsel:2007if,Tiator:2008kd}. In particular, due to the isosinglet nature of the involved electromagnetic current, the usually dominant contribution from the $\Delta$-resonances is absent in $F_3^{(0)}$, which makes the entire resonance contribution small. 

\begin{figure}[t]
\begin{center}
\includegraphics[width=0.3\columnwidth]{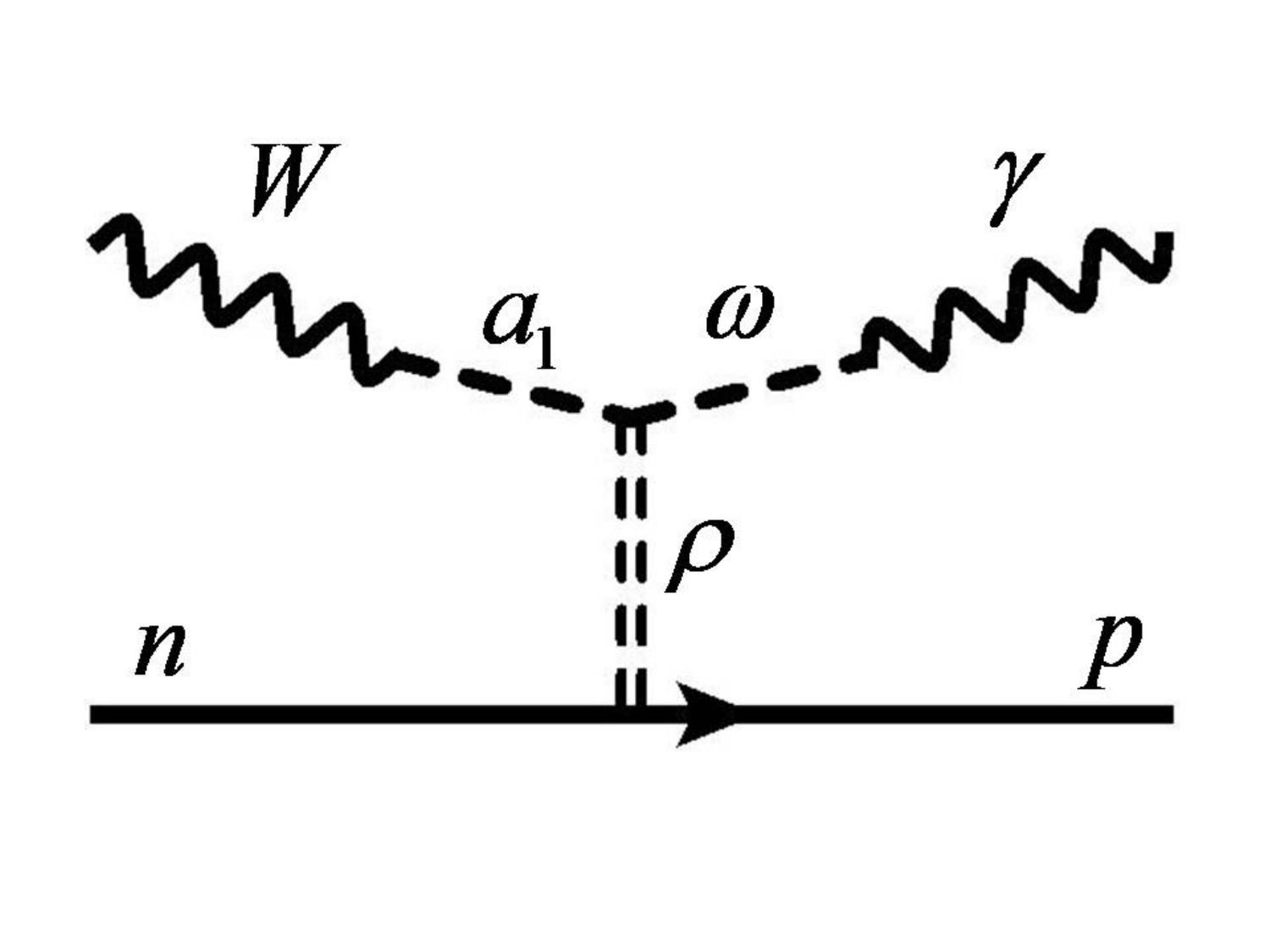}
\includegraphics[width=0.3\columnwidth]{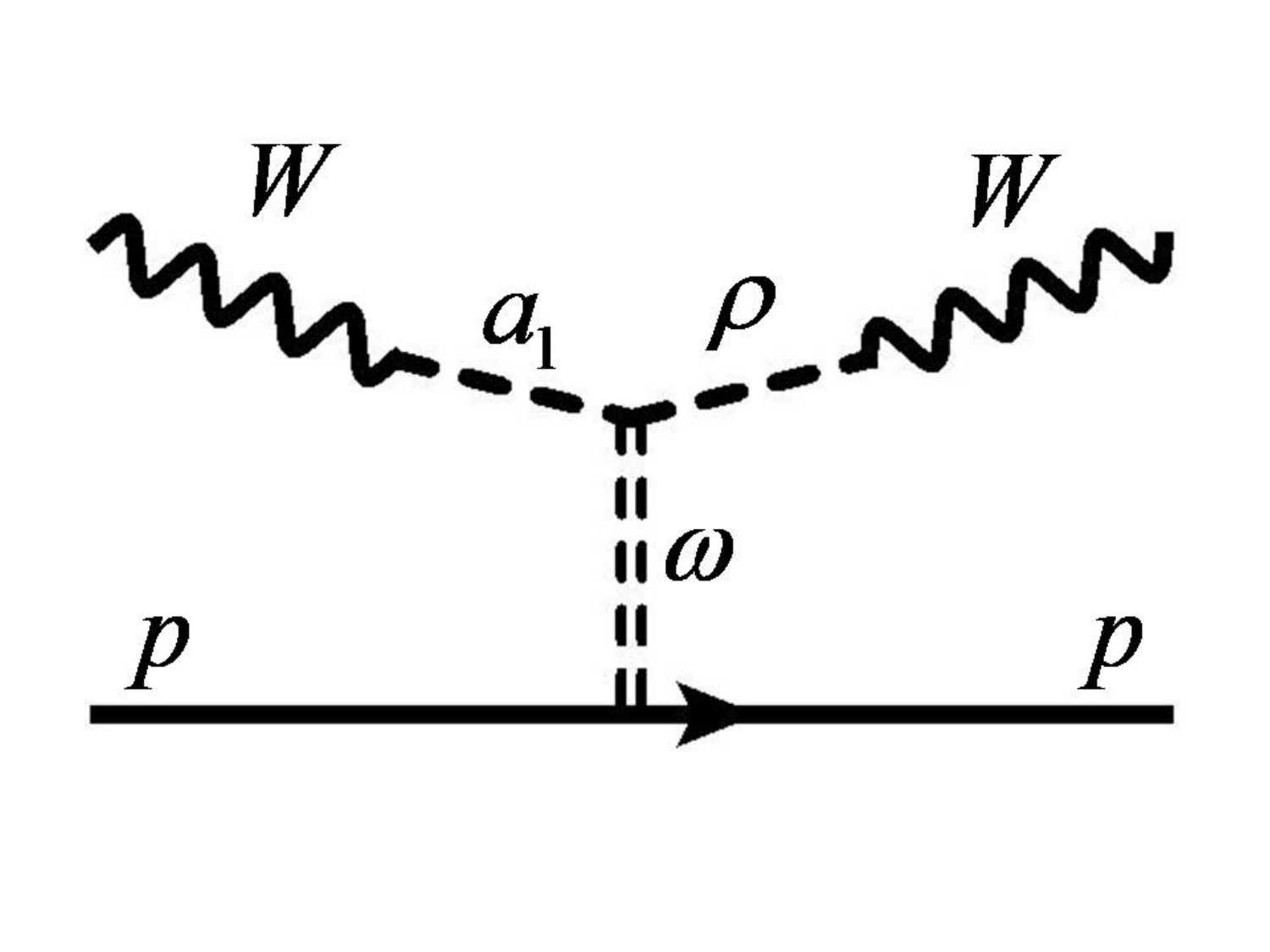}
\includegraphics[width=0.3\columnwidth]{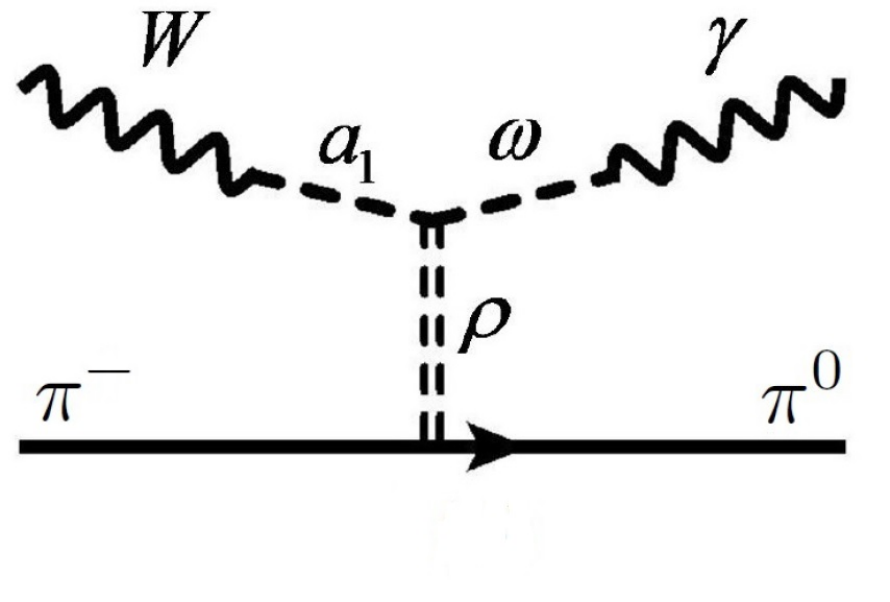}
\caption{Leading Regge contribution to $F_{3,N}^{(0)}$, $F_3^{\nu p+\bar{\nu}p}$ and $F_{3,\pi}^{(0)}$ respectively.}
\label{fig:Regge}
\end{center}
\end{figure}

The main difference in the existing literature resides in the treatment of the subasymptotic high-energy but low-$Q^2$ contribution. In the dispersive representation adopted in Refs.~\cite{Seng:2018qru,Seng:2018yzq,Shiells:2020fqp,Seng:2020wjq}, a natural language for the contributions from this kinematical range is that of $t$-channel Regge exchanges supplemented by the (axial) vector dominance model (VDM). For $F_3^{(0)}$, this contribution is depicted by the first diagram in Fig.~\ref{fig:Regge}. The same qualitative picture applies to the structure functions in other reactions, e.g. $F_3^{\nu p+\bar{\nu}p}$ in inclusive $\nu p/\bar{\nu}p$ scattering~\cite{Bolognese:1982zd,Kataev:1994rj,Kim:1998kia,Allasia:1985hw}, described by the second diagram in Fig.\ref{fig:Regge}. Given the near degeneracy of the $\rho$- and $\omega$-trajectories, these two diagrams only differ by the gauge boson-meson coupling and meson-nucleon coupling constants. The former follows a simple scaling in the VDM picture \cite{Lichard:1997ya}, while the latter follows from the straightforward isospin scaling in the Regge-exchange picture. This leads to an approximate scaling of the first Nachtmann moments,
\begin{equation}
    M_{3,\mathbb{R}}^{(0)}(1,Q^2)\approx \frac{1}{18}M_{3,\mathbb{R}}^{\nu p+\bar{\nu}p}(1,Q^2)~.\label{eq:matching}
\end{equation}
This allows one to constrain the problematic subasymptotic piece with the use of inclusive data from neutrino/antineutrino scattering experiments. 
%We emphasize here that the scaling relation of Eq.\eqref{eq:matching} is based on the assumption of the approximate validity of the VDM picture and Regge behavior of the high-energy cross sections.
This strategy was adopted by Refs.~\cite{Seng:2018qru,Seng:2018yzq}. Ref.\cite{Shiells:2020fqp} did a more careful analysis of the correspondence between $F_3^{(0)}$ and $F_3^{\nu p+\bar{\nu}p}$ as a function of $x_B$ and $Q^2$, and supported the Regge-VDM Ansatz of  Eq.\eqref{eq:matching}.

Another source of input that allows to constrain the Regge contribution comes from recent lattice QCD calculations of the pion $\gamma W$-box diagram~\cite{Feng:2020zdc,Yoo:2022lmt} which will be described in section \ref{sec:LQCD}. One fundamental property of Regge exchanges is the factorizability, well-established both theoretically and experimentally. Factorizability states that, e.g. a $t$-channel Regge $\rho$-exchange contribution to a generic $ab\to cd$ scattering amplitude factorizes as  
\begin{align}
    T_{ab\to cd}=\Gamma_{ac}(t)P_\rho(s,t)\Gamma_{bd}(t)\,,
\end{align}
with vertex functions $\Gamma_{ac}$ being specific for a particular process, and the Regge propagator $P_\rho(s,t)$ being universal for any process where the $\rho$ quantum numbers can be exchanged. This relates the ratio of the electroweak amplitudes enering the $\gamma W$-box calculation to that of purely  hadronic amplitudes:
\begin{align}
\frac{F_{3,\pi,\mathbb{R}}^{(0)}}{F_{3,N,\mathbb{R}}^{(0)}}
=\frac{T_{\pi\pi\to\pi\pi}^\rho}{T_{\pi N\to\pi N}^\rho}=\frac{T_{\pi N\to\pi N}^\rho}{T_{NN\to NN}^\rho}\,.\label{eq:Reggefact}
\end{align}
These latter ratios are known from a partial wave analysis of high-energy $\pi\pi$, $\pi N$ and $NN$ elastic scattering \cite{Caprini:2011ky}. Note that that analysis included the test of Regge factorization, the second equality in Eq.~\eqref{eq:Reggefact}.

To obtain the Regge contribution to $F_{3,\pi}^{(0)}$, depicted by the third diagram in Fig.~\ref{fig:Regge}, from the full lattice result, the non-asymptotic low-energy part, specific to the pion decay, has to be removed. This latter contribution comprises $2\pi,4\pi,\dots$ intermediate states in the $s$-channel, and is dominated by a single $\rho$ resonance. It turns out to be numerically small \cite{Seng:2020wjq}, thus not introducing a large systematic uncertainty. With this approach one could infer  the subasymptotic Regge part of the neutron $\gamma W$-box from the lattice calculation of its pion counterpart.

Finally, the properly asymptotic piece follows from the parton model and pQCD and is the same in all approaches, modulo small subleading corrections. It largely coincides with that to $\Box_{\gamma W}^A$ reviewed above in great detail.

\subsection{Non-dispersive determination of $\Box_{\gamma W}^V$}

As a comparison, we also review an approach based on a non-dispersive formalism, pioneered by Marciano and Sirlin in 2006~\cite{Marciano:2005ec} and improved upon by the same authors and Czarnecki~\cite{Czarnecki:2019mwq}. The vector box diagram is expressed as
\begin{equation}
    \Box_{\gamma W}^V=\frac{\alpha_\text{em}}{8\pi}\int_0^\infty dQ^2\frac{M_W^2}{M_W^2+Q^2}F(Q^2)~,
\end{equation}
with the straightforward correspondence  $F(Q^2)=(6/Q^2)M_3^{(0)}(1,Q^2)$. 
At asymptotically large $Q^2$, $F(Q^2)$ takes the form of a Mellin moment, and behaves as~\cite{Marciano:2005ec}
\begin{equation}
    F(Q^2)\rightarrow F_\text{pQCD}(Q^2)=\frac{1}{Q^2}\left(1-\frac{\alpha_{g_1}(Q^2)}{\pi}\right)~.
\end{equation}
The terms in the round bracket represent the pQCD correction identical to that in the isovector BjSR for the spin-dependent structure function $g_1$~\cite{Bjorken:1966jh,Bjorken:1969mm}:
\begin{equation}
\int_0^1dx_B[g_1^p(x_B,Q^2)-g_1^n(x_B,Q^2)]=\frac{|\mathring{g}_A|}{6}\left(1-\frac{\alpha_{g_1}(Q^2)}{\pi}\right)~,~\text{large}~Q^2~.\label{eq:Fasymp}
\end{equation}
This pQCD correction is known to order $\alpha_s^4$~\cite{Baikov:2010je}; the same treatment was also adopted in the DR-based work. 

The revised treatment of Ref.~\cite{Czarnecki:2019mwq} extends the discussion based on BjSR to the non-perturbative domain for which they proposed two parameterizations of $F(Q^2)$ at small $Q^2$. The first one is based on the AdS light-front holographic QCD (LFHQCD)~\cite{Brodsky:2003px,Brodsky:2014yha} which offers a good qualitative description of the  experimental data on BjSR at low $Q^2$~\cite{Deur:2014vea}. It entails using Eq.\eqref{eq:Fasymp} below the perturbative region, but with 
\begin{equation}
    \frac{\alpha_{g_1}(Q^2)}{\pi}=\exp(-Q^2/Q^2_\text{AdS})~,~Q^2<Q_\text{AdS}^2\,,
\end{equation}
where $Q_\text{AdS}^2=1.10(10)$~GeV$^2$ is the matching scale between the perturbative and non-perturbative regimes in this prescription, determined by the continuity of $F(Q^2)$ at $Q^2=Q_\text{AdS}^2$. The second parameterization is a three-resonance form,
\begin{equation}
    F_\text{res}(Q^2)=\frac{A}{Q^2+m_\rho^2}+\frac{B}{Q^2+m_A^2}+\frac{C}{Q^2+m_{\rho'}^2}~,~Q^2<Q_\text{AdS}^2
\end{equation}
similar to their previous work, Ref.\cite{Marciano:2005ec}, but here $Q_\text{AdS}^2$ is taken from the AdS prediction above. The three coefficients $A,B,C$ are fixed by the three conditions,
\begin{enumerate}
    \item The integral $\int_{Q_\text{AdS}^2}^\infty dQ^2M_W^2 F(Q^2)/(M_W^2+Q^2)$ is equal for $F=F_\text{pQCD}$ and $F=F_\text{res}$.
    \item The $1/Q^4$ term in $F_\text{res}(Q^2)$ is required to vanish at large $Q^2$. 
    \item 
    %The value of the integrand at $Q^2=Q_\text{AdS}^2$ is taken to be same as the AdS prediction,
    $F_\text{res}(Q_\text{AdS}^2)=F_\text{AdS}(Q_\text{AdS}^2)$. 
\end{enumerate}
Both parameterizations returned consistent results for $\Box_{\gamma W}^V$, which are larger than the 2006 determination but smaller than the DR determination.

\subsection{Recommended values for $\Box_{\gamma W}^V$, $\Delta_R^V$, $\Delta_R$, $V_{ud}$}
\label{sec:RecommendedValues}

\begin{table}[h]
\begin{centering}
\begin{tabular}{|c|c|c|c|c|c|c|}
\hline 
& \cite{Seng:2018qru,Seng:2018yzq} & \cite{Seng:2020wjq}  & \cite{Shiells:2020fqp} & \cite{Czarnecki:2019mwq} &
\cite{Hayen:2020cxh}
 & Our value\tabularnewline
\hline 
\hline 
Born  & 1.06(6) & 1.06(6) & 1.05(4) & 0.99(10) & 1.06(6) & 1.06(6)\tabularnewline
\hline 
$\pi N+$Res. & 0.05(1) & 0.05(1)  & 0.04(1) & - & - & 0.05(1)\tabularnewline
\hline 
Regge & 0.51(8) & 0.56(9) & 0.52(7) & 0.38(3) & 0.53(7) & 0.54(6)\tabularnewline
\hline 
DIS & 2.17 & 2.16 & 2.20(3) & 2.16 & 2.16 & 2.20(3)\tabularnewline
\hline 
\end{tabular}
\par\end{centering}
\caption{\label{tab:compare}Various contributions to $\Box_{\gamma W}^V$, in units of $10^{-3}$, in existing literature. The first three columns are DR-based works, the following two are non-DR-based. Regge or Interpolator contribution is understood for the DR and non-DR approaches, respectively. The last column quotes our recommended values. 
The table entries differ from Table 2 in Ref.\cite{Cirigliano:2022yyo}, see the explanations in Sec.\ref{sec:RecommendedValues}.}
\end{table}

A detailed comparison of the outcome of different approaches is given in Table~\ref{tab:compare}
separately for the Born, $\pi N+$resonance, Regge (the LFHQCD/three-resonances interpolator for the non-DR approaches) and DIS. A similar table was presented in Ref.~\cite{Cirigliano:2022yyo}, but some of its entries are different from ours, which we explain as follows:
\begin{itemize}
\item Both tables define the ``DIS'' entry as everything above $Q^2=Q_0^2=2$~GeV$^2$; but the non-DR papers adopts $Q_\text{AdS}^2\approx 1.10$~GeV$^2$ instead as the separation scale between the ``perturbative'' and ``non-perturbative'' region of their integral. In order to translate the latter, Ref.\cite{Cirigliano:2022yyo} subtracts from the DIS results in the non-DR papers an estimated value of the ``DIS contribution from 1 to 2~GeV$^2$'' given in Ref.\cite{Shiells:2020fqp}. In this review we do not follow such a prescription, but compute instead the $Q^2>2$~GeV$^2$ integral directly using the known analytic formula from pQCD. 
\item Authors in Ref.\cite{Cirigliano:2022yyo} followed Ref.\cite{Shiells:2020fqp}and include the effect of the running $\alpha_\text{em}$ in the box diagram; to do so they need to manually increase the DIS results in Refs.\cite{Seng:2018qru,Seng:2018yzq,Seng:2020wjq,Czarnecki:2019mwq,Hayen:2020cxh} by about 4\%; on the contrary, in this work we define $\Box_{\gamma W}$ with a constant $\alpha_\text{em}$ and move the running effect into $\delta_\text{HO}^\text{QED}$ following Ref.\cite{Czarnecki:2004cw}. Consequently, our DIS result is lower than that in Ref.\cite{Shiells:2020fqp} by 4\%. 
\item 
We display the uncertainty of the ``Regge'' part of the non-DR works, which was omitted in
Ref.\cite{Cirigliano:2022yyo}.
\end{itemize}

The exact value of the neutron $\gamma W$-box is central to extracting $V_{ud}$ from various beta decay measurements, and its uncertainty is crucial for defining the CKM unitarity deficit. 
Since one observes a certain spread in the individual numbers, it is worth devising a reliable average value and its uncertainty. For this, we will consider each row separately and discuss the reason for the differences in both central value and the respective uncertainty. 

For the Born contribution there is a great consistency between all central values apart from that of Ref.~\cite{Czarnecki:2019mwq} which is on a lower side. The reason for this is well-known: the authors of \cite{Czarnecki:2019mwq} only integrate the elastic contribution up to a low $Q^2=Q_\text{AdS}^2\approx1.10$ GeV$^2$, whereas all other references integrate to infinity. From the dispersive perspective there is no ambiguity: the elastic contribution is present at any value of $Q^2$ as an isolated pole at $W^2=M^2$ separated from inelastic contributions by a finite energy gap. One must therefore integrate over all values of $Q^2$. We conclude that the low value for $\Box_{\gamma W}^{V\,,\mathrm{Born}}$ from Ref.~\cite{Czarnecki:2019mwq} has to be dropped. Since all groups use the same form factor data, no averaging is needed, and we quote the conservative 
\begin{align}
\overline{\Box}_{\gamma W}^{V\,,\mathrm{Born}}=1.06(6)\times 10^{-3}. 
\end{align}
The low-energy $\pi N$ and resonance contribution appears only in the DR-based approaches (the non-DR works lump everything into a single ``non-perturbative'' contribution, which we label as ``Regge'' in Table~\ref{tab:compare}) that use the same ingredients. We take therefore 
\begin{align}
\overline{\Box}_{\gamma W}^{V\,,\pi N+\mathrm{Res}}=0.05(1)\times 10^{-3}\,,
\end{align}
again, without the need to average. %since the starting point is the PBjSR that concerns the spin-dependent structure functions: the low-energy excitations in $F_3^{(0)}$ and $g_1^{(0)}$ are unrelated. 

The third row is where the bulk of the discrepancy between the non-dispersive and dispersive evaluations  resides. First, we address the non-DR based subasymptotic contribution. 
As explained in the previous subsection, it is based on the BjSR which operates with the first Mellin moment, while the $\Box_{\gamma W}^V$  integrand involves the Nachtmann moment instead. The two become close only at large $Q^2$, but at small $Q^2$ they differ by the target-mass corrections (TMC) that are not negligible. This correction was included in the AdS parameterization by Hayen~\cite{Hayen:2020cxh}, leading to a shift upward. Note that TMC are automatically included in the DR-based approach. 

Furthermore, a rigorous quantification of the uncertainty from the matching of the non-perturbative LFHQCD model to pQCD is challenging. As seen in Fig.~\ref{fig:Gammabar1}, the higher-twist contribution and the respective uncertainty is non-negligible already at 2\,GeV$^2$. In directly applying the subasymptotic part derived from the BjSR to the first Mellin moment of $F_3^{(0)}$, 
Ref.~\cite{Czarnecki:2019mwq} assumes no HT contribution, and no associated uncertainty. 
%As well, it misses the non-asymptotic resonance contribution which is not captured by the analogy with the BjSR.
Ref.~\cite{Hayen:2020cxh} includes the HT correction to the asymptotic part, but largely underestimates its size and uncertainty for BjSR, as already mentioned. 
Furthermore, Ref.~\cite{Hayen:2020cxh} 
observed that the HT contribution for $M_3^{(0)}$ has a positive sign, opposite to that in BjSR. This observation is compatible with the lattice-driven estimate of Ref.~\cite{Seng:2020wjq},
albeit seems to underestimate its size. 
%which indicates that in the case of $M_3^{(0)}$ the subasymptotic correction approaches the perturbative prediction from above (see Fig.~2 of that reference). 
%However, as in the case of BjSR, Ref.~\cite{Hayen:2020cxh} 
%but lattice QCD calculation indicates a larger size
Given that matching of the subasymptotic part in non-DR approaches occurs at a low $Q^2\sim1$ GeV$^2$, this results in a systematic underestimation of the subleading contribution and its uncertainty. 

We conclude that both non-DR estimates of the subasymptotic piece contain uncontrolled systematic uncertainties. Most importantly, the true uncertainty of the Regge contribution of Ref.~\cite{Czarnecki:2019mwq} in Table~\ref{tab:compare} has to be significantly asymmetric towards larger values.
Therefore we deem that the non-DR estimates of the Regge contribution, while generally lending qualitative support to the dispersive evaluations, should not be taken into account in the average. Of the remaining DR-based results, those of Refs.~\cite{Seng:2018qru,Seng:2018yzq,Shiells:2020fqp} are based on the same low-$Q^2$ $\nu/\bar\nu$ DIS data, whereas that of Ref.~\cite{Seng:2020wjq} is based on the input from lattice QCD. 
Our average thus is obtained from the most precise data-driven result of Ref.~\cite{Shiells:2020fqp} and the lattice QCD-driven one of Ref.~\cite{Seng:2020wjq}, and reads:
 \begin{align}
\overline{\Box}_{\gamma W}^{V\,,\mathrm{Regge}}=0.54(6)\times 10^{-3}. 
\end{align}

Finally, we address the DIS contribution.  The starting point of all calculations is the 4-loop BjSR expression in Eq.~\eqref{eq:BjSRpQCD} which is integrated from the matching point $Q_0^2$ to $\infty$. Among them, only Ref.\cite{Shiells:2020fqp} carefully studied the high-$Q^2$ contribution by making use of the available parameterizations of parton distribution functions beyond their sum rules. This approach leads to a more robust theory uncertainty, and we adopt their number: 
\begin{align}
\overline{\Box}_{\gamma W}^{V\,,\mathrm{DIS}}=2.20(3)\times 10^{-3}. 
\end{align}
%{\color{red} Chien Yeah, please complete this part. Discuss Hayen number being higher? Is my explanation to Shiells number correct?} 
%Finally, it has been proposed to include the running of $\alpha$ in the DIS part of the $\gamma W$-box. To do so one has to remove the respective effect from the resummed leading logarithms at order $O(\alpha^2)$ \cite{Czarnecki:2004cw,Cirigliano:2022yyo} 
%$S(M_p,M_Z)\to\tilde S(M_p,M_Z)$ 
%to avoid double-counting. Note that
%this adjustment was not done in Ref.~\cite{Shiells:2020fqp}. We opt not to include the running of the electromagnetic constant into the $\gamma W$-box but keep the original HO correction of Ref.~\cite{Czarnecki:2004cw}. In doing so, we wish to avoid further confusion, keeping in mind that the resummation performed in Ref.~\cite{Czarnecki:2004cw} only accounts for a part of the HO leading logs, and emphasize the need in a complete calculation.
%
Adding the ``universal'' contribution $\Delta_R^U$ in Eq.\eqref{eq:universal}, our final recommended values are thus 
\begin{align}
\overline{\Box}_{\gamma W}^V=3.85(9)\times10^{-3},\quad\Delta_R^V=0.02479(21),\quad \Delta_R=0.03985(21),%,\quad\Delta_R=\dots.
\end{align}
%{\color{red} Chien Yeah, please check and complete} 
where $\Delta_R\equiv \Delta_R^V+(\alpha_\text{em}/2\pi)\bar{\delta}^{(1)}$ is a customarily-defined quantity for the ``full'' RC to neutron lifetime. Finally, employing these results, we obtain the following value of $V_{ud}$ from neutron decay with the PDG-averaged and best-value $\{\tau_n,\lambda\}$, respectively:
\begin{equation}
 |V_{ud}|_n^{\rm PDG-av}=0.97433(28)_{\tau_n}(82)_\lambda(10)_\text{RC}~,\quad
  |V_{ud}|_n^{\rm best}=0.97404(20)_{\tau_n}(35)_\lambda(10)_\text{RC}~.
\end{equation}
We note that both extractions are significantly higher than that from the superallowed nuclear decays \cite{ParticleDataGroup:2022pth}, $|V_{ud}|_{0^+}=0.97367(31))$. These results can further be combined with $V_{us}$ from kaon decays to set constraints on the CKM unitarity violation in the top row (dropping the small $V_{ub}$ contribution), $\Delta_u =|V_{ud}|^2+|V_{us}^2|-1$.
The current PDG average  \cite{ParticleDataGroup:2022pth} reads $|V_{us}|=0.2243(8)$, and
%with the subscripts indicating the semileptonic decay channels $K\ell2=K\to\ell\nu_\ell$ combined with $\pi\ell2$, and $K\ell3=K\to\pi\ell\nu_\ell$, respectively. 
we obtain 
\begin{align}
\Delta_u^{\text{n,\,PDG-av}}=-0.00037(174)\,,\quad 
\Delta_u^{\text{n,\,best}}=-0.00094(89)\,,
\end{align}
showing no statistically significant deviation from unitarity. Given the 2.5$\sigma$ deficit observed if the nuclear beta decays are used, $\Delta_u^\text{nuclear}=-0.00166(69)$, future improvement of the experimental precision in neutron decay become even more important.

\section{Lattice QCD}
\label{sec:LQCD}
An important step forward in reducing the hadron structure uncertainties in neutron beta decay is to rely on lattice QCD. 
To further improve upon existing work on the free neutron RC, a direct lattice QCD computation is required. Unlike computations of bare QCD coupling constants (e.g. $\mathring{g}_A$) which have been going on for almost two decades~\cite{FlavourLatticeAveragingGroupFLAG:2021npn}, the lattice study of RC in neutron decay is a relatively new subject, which we will briefly review in this section.  

\subsection{The $\gamma W$-box diagram: semileptonic pion and kaon decays}

To study the box diagram corrections $\Box_{\gamma W}^{V,A}$ from first principles requires one to compute the invariant amplitudes $T_3$ and $g_{1,2}$ from lattice QCD at low $Q^2$. At high $Q^2$, large lattice artefacts make such calculations unreliable,  but one can compute these amplitudes with pQCD and  combine lattice calculation at low $Q^2$ with pQCD calculation at high $Q^2$. 

\begin{figure}[h]
	\begin{centering}
		\includegraphics[scale=0.5]{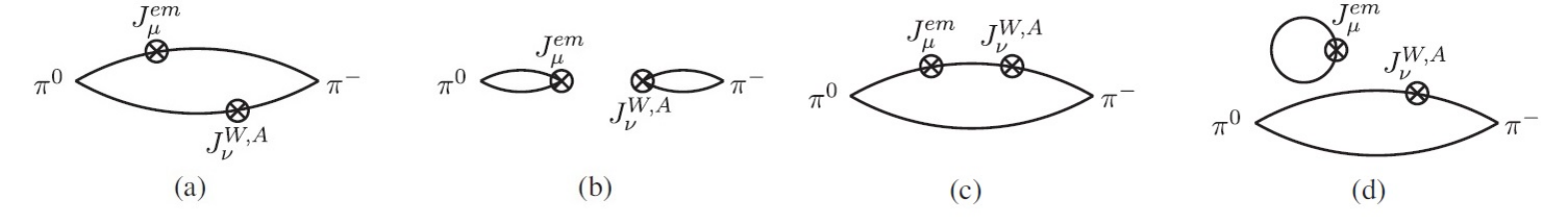}\hfill
		\par\end{centering}
	\caption{\label{fig:4pt}Quark contraction diagrams that correspond to $\mathcal{H}_{\mu\nu}^{VA}(x)$. Figures are reproduced from Ref.~\protect\cite{Feng:2020zdc}.}
\end{figure}

The first attempt was carried out by the RBC-UKQCD Collaboration in Ref.\cite{Feng:2020zdc}. They worked on a simpler case, namely the box diagram in the $\pi^-\rightarrow \pi^0$ decay to demonstrate the applicability of their method. At low $Q^2$, the following Euclidean spacetime integral was computed:
\begin{equation}
M_\pi(Q^2)=-\frac{1}{6\sqrt{2}}\frac{\sqrt{Q^2}}{m_\pi}\int d^4x\omega(Q,x)\epsilon_{\mu\nu\alpha 0}x_\alpha \mathcal{H}_{\mu\nu}^{VA}(x)~,\label{eq:MpiQ2}
\end{equation}
where $\omega(Q,x)$ is a known function, and 
\begin{equation}
\mathcal{H}_{\mu\nu}^{VA}(x)=\langle \pi^0(P)|T[J_\mu^\mathrm{em}(x)J_\nu^{W,A}(0)]|\pi^-(P)\rangle
\end{equation}
is a four-point correlation function consists of quark contraction diagrams depicted in Fig.\ref{fig:4pt}. They found that the lattice result joins smoothly to the pQCD prediction at $Q^2\approx 2$~GeV$^2$, lending support to the previous choice of the separation scale $Q_0^2$. With this prescription, they obtained a prediction of $\Box_{\gamma W,\pi}^V$ at percent level. Following the suggestion in Refs.\cite{Seng:2019lxf,Seng:2020jtz}, the same group performed a second calculation that corresponds to the box diagram correction in the $\bar{K}^0\rightarrow \pi^+$ decay in the flavor SU(3) limit~\cite{Ma:2021azh}, which helped to fix a number of important LECs responsible for semileptonic decays of pions and kaons. 

\begin{table}
\begin{centering}
\begin{tabular}{|c|c|c|}
\hline 
 & RBC-UKQCD & LANL\tabularnewline
\hline 
\hline 
$\Box_{\gamma W,\pi}^{V}$ & $2.830(28)\times10^{-3}$ & $2.810(26)\times10^{-3}$\tabularnewline
\hline 
$\Box_{\gamma W,K}^{V}$ & $2.437(44)\times10^{-3}$ & $2.389(17)\times10^{-3}$\tabularnewline
\hline 
\end{tabular}
\par\end{centering}
\caption{\label{tab:comparelat}Comparison between the RBC-UKQCD~\cite{Feng:2020zdc,Ma:2021azh} and LANL~\cite{Yoo:2023gln} result of the pion and kaon box diagram calculations. The $\sim 1\sigma$ difference between the two calculations of $\Box_{\gamma W,K}^V$ is mainly due to the difference in the choice of the SU(3)-symmetric point.}

\end{table}

The LANL lattice group performed an independent calculation of the two quantities above~\cite{Yoo:2023gln}, and found good agreement with the RBC-UKQCD result, see Table~\ref{tab:comparelat}. Both groups are currently working on the $\Box_{\gamma W}^V$ for neutron. It will be very instructive to compare it, once available, to the hybrid lattice QCD - phenomenology result of Ref.~\cite{Seng:2020wjq}.

A similar calculation of $\Box_{\gamma W}^A$ can also be performed. Given that the latter is very precisely determined in the DR framework using experimental data of $g_{1,2}$, it may serve as an important cross-check of the lattice accuracy. It is also worth to notice that, apart from the direction that entails computing quark contraction diagrams, alternative approaches have been proposed to study the structure functions in the box diagrams. One such approach is based on the Feynman-Hellmann theorem~\cite{Seng:2019plg}.
\subsection{RC to the nucleon axial coupling constant}

As shown in Ref.\cite{Cirigliano:2022hob} which we will discuss in the next section, the $\gamma W$-box diagram is not the only non-trivial component in RC when it comes to the axial coupling constant $g_A$; a much larger effect comes from the form factor correction, i.e. the left diagram in Fig.\ref{fig:NPRC}. Therefore, a direct  lattice calculation is desirable. The expression of $\delta F^\mu$ in Sirlin's representation is not the most convenient starting point for this purpose, because the three-point function $\delta F_\text{3pt}^\mu$ is, in the language of lattice QCD, a ``five-point function'' (two external states and three current insertions) which is extremely difficult to handle. 

A more tractable alternative is to directly compute the single-nucleon matrix element $\langle p|J_W^\mu|n\rangle$ in the presence of electromagnetic interactions. A way to proceed is to adopt lattice QCD + Quantum Electrodynamics (QED) with massive photons~\cite{Endres:2015gda}, also referred to as QED$_\text{M}$. Given that $\delta F^\mu$ contains infrared-divergences, one requires also an analytical expression of the terms that depend on the fictitious photon mass $m_\gamma$. Finally, while $\delta F_\text{3pt}^\mu$ is complicated, the two-point function $\delta F_\text{2pt}^\mu$ takes a much simpler form and may be directly calculable on lattice using its definition in Eq.\eqref{eq:2pt}. A separate lattice calculation of the full $\delta F^\mu$ (using QED$_\text{M}$) and $\delta F_\text{2pt}^\mu$ (by computing the quark contraction diagrams) would allow for a comparison of the outcomes to their respective low-energy theory expression. 

%{\color{red} Maybe a summary table of various calculations of $\Delta_R^V$, $\Delta_R^A$, $V_{ud}$ and $g_A$ would be handy?}
%%%%%%%%%%%%%%%%%%%%%%%%%%%
\section{Effective field theory description of radiative corrections\label{sec:EFT}}

In parallel to the aforementioned studies based on Sirlin's representation of RC, there are also attempts to study neutron RC in an effective field theory (EFT) language, which we briefly overview in this section. Some advantages of this approach are that calculations may proceed using standard Feynman rules, theory precision is systematically improvable following well-defined power counting rules, and sources of uncertainties clearly identified in terms of low-energy constants (LECs) in the theory. First attempt of such kind was based on a pionless effective Lagrangian expanded to  $\mathcal{O}(1/M)$~\cite{Ando:2004rk,Bernard:2004cm}: 
\begin{equation}
\mathcal{L}=\mathcal{L}_{e\nu\gamma}+\mathcal{L}_{NN\gamma}+\mathcal{L}_{e\nu NN}~,
\end{equation}
where explicitly form of respective terms can be found in Ref.\cite{Ando:2004rk}. One-loop calculations with the Lagrangian above reproduces known expressions of Fermi function (to order $\mathcal{O}(\alpha_\text{em})$), outer corrections and recoil corrections to the differential decay rate. 

A problem of the pionless EFT is that it has no predictive power on the \textit{inner} corrections, which are entirely encapsuled in the (already) renormalized coupling constants and LECs that appears in the Lagrangian. 
A recent re-evaluation by Cirigliano \textit{et al.}~\cite{Cirigliano:2022hob} made use of the heavy baryon chiral perturbation theory (HBChPT), which includes pions as dynamical degrees of freedom. A power counting rule $p/\Lambda_\chi\sim m_\pi/\Lambda_\chi\sim e$ is adopted, where $\Lambda_\chi\approx 1$~GeV is the chiral symmetry breaking scale. With this, a chiral order is assigned to every term in the chiral Lagrangian:
\begin{eqnarray}
\mathcal{L}_\pi&=&\mathcal{L}_\pi^{(2)}+...\nonumber\\
\mathcal{L}_{\pi N}&=&\mathcal{L}_{\pi N}^{(1)}+\mathcal{L}_{\pi N}^{(2)}+\mathcal{L}_{\pi N}^{(3)}+...\nonumber\\
\mathcal{L}_\text{lept}&=&\mathcal{L}_\text{lept}^{(1)}+\mathcal{L}_\text{lept}^{(2)}+...
\end{eqnarray}
With this formalism one could predict, in addition to the outer corrections and recoil corrections, a part of the inner corrections that are associated to pion loops. 

\begin{figure}
	\begin{centering}
		\includegraphics[scale=0.6]{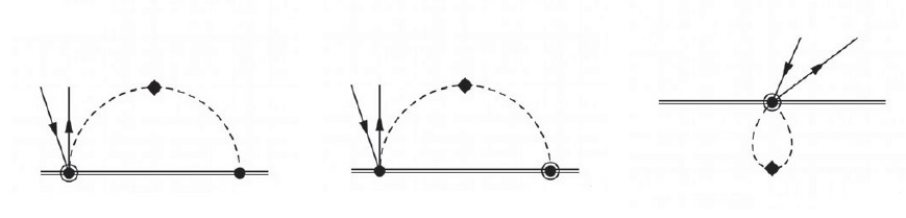}\hfill
		\par\end{centering}
	\caption{\label{fig:NLO}Next-to-leading-order HBChPT diagrams that contribute to the neutron axial form factor. Dots, diamonds and circled dots represent vertices derived from the $\mathcal{O}(p)$, $\mathcal{O}(e^2)$ and $\mathcal{O}(p^2)$ chiral Lagrangian respectively. Figure courtesy of V. Cirigliano. }
\end{figure}

The most important result in Ref.\cite{Cirigliano:2022hob} is the renormalization of the axial coupling $g_A$, which they expressed as:
\begin{equation}
    g_A=\mathring{g}_A\left[1+\sum_{n=2}^\infty\Delta_{A,\chi}^{(n)}+\frac{\alpha_\text{em}}{2\pi}\sum_{n=0}^\infty\Delta_{A,\text{em}}^{(n)}+\left(\frac{m_u-m_d}{\Lambda_\chi}\right)\sum_{n=0}^\infty \Delta_{A,\delta m}^{(n)}\right]~,
\end{equation}
where the three terms at the RHS correspond to the QCD correction away from the chiral limit, the electromagnetic (EM) correction and the strong ISB correction, respectively, and they concentrated on the EM correction $\Delta_{A,\text{em}}^{(n)}$. They found that, the leading-order (i.e. $\mathcal{O}(e^2)$) correction to $g_A$ depends on unknown LECs to cancel the ultraviolet-divergences in the loop diagrams, but the next-to-leading-order (i.e. $\mathcal{O}(e^2p)$) correction is fully predictable. The latter are given by the Feynman diagrams in Fig.\ref{fig:NLO}. Notice that these diagrams correspond to the left diagram in Fig.\ref{fig:NPRC} and are independent of the $\gamma W$-box diagram that we discussed in Sec.\ref{sec:gammaW}. The outcome reads:
\begin{equation}
\Delta_{A,\text{em}}^{(1)}=4\pi m_\pi Z_\pi\left[c_4-c_3+\frac{3}{8M}+\frac{9}{16M}\mathring{g}_A^2\right]~,
\end{equation}
where $Z_\pi$ is determined from the $m_{\pi^+}-m_{\pi^0}$ mass splitting, and $c_{3,4}$ are obtained from pion-nucleon scattering. With existing fits of LECs they obtained $\Delta_{A,\text{em}}^{(1)}=10.0$--$15.9$, namely the inner RC to $g_A$ is of the order $10^{-2}$. This finding has no consequence in the extraction of $V_{ud}$ from neutron beta decay because the experimental value of $\lambda$ is fully renormalized, but it has a huge impact if one is to compare the experimental $g_A$ and the bare $\mathring{g}_A$, the latter obtained from lattice QCD calculations, in order to constrain new physics. 

It could be useful to connect the EFT result to Sirlin's representation. As we discussed above, $\Delta_{A,\text{em}}^{(1)}$ which originates from diagrams in Fig.\ref{fig:NLO}, belong to the RC to the charged weak current matrix element, i.e. $\delta F^\mu=\delta F_\text{2pt}^\mu+\delta F_\text{3pt}^\mu$ in Eq.\eqref{eq:Fmusplit}. In fact, it must come from the three-point function because it only involves short-distance electromagnetic effects (characterized by the LEC $Z_\pi$) which are not present in the two-point function. The largeness of $\Delta_{A,\text{em}}^{(1)}$ proves our assertion at the end of Sec.\ref{sec:inner}. In fact, it will be a useful exercise to compute just $\delta F^\mu$, instead of the full RC, using HBChPT, and split it into two-point and three-point functions using their definitions in Eqs.\eqref{eq:2pt}, \eqref{eq:3pt} respectively. This will benefit the future lattice study of $g_A$ RC which we describe in the previous section. 

Finally, during the preparation of this manuscript, a more complete top-down EFT analysis of the SM RC appeared~\cite{Cirigliano:2023fnz}. Two main results of the paper are: (1) It claimed to have corrected a mistake in the calculation of $\delta_\text{HO}^\text{QED}$, and (2) The ``conventional'' Fermi function was replaced by a ``non-relativistic'' version independent of the proton charge radius, which affects the statistical rate function $f$:
\begin{equation}
\delta \Delta_R^U\approx +0.061\%~,~\delta f\approx -0.035\%~.
\end{equation}
The two changes are of different sign and partially compensate, leading to a slight positive shift of $+0.026\%$ of the neutron decay rate, which corresponds to a negative shift in $V_{ud}$ from neutron decay, $\delta V_{ud}\approx -1.3\times 10^{-4}$. It is important to cross-check these new results and discuss their implications for more general charged-weak decay processes. 

%%%%%%%%%%%%%%%
\section{Searches for physics beyond the Standard Model: beta decays vs. LHC}\label{sec:BSM}

With the experimental and theoretical precision reaching few parts in $10^4$, neutron decay is a promising avenue to look for deviations from the SM predictions due to the presence of new particles and/or interactions. Heavy new particles at the TeV scale may become visible at the Large Hadron Collider (LHC), but no deviations have been unambiguously detected in the currently accessible energy range. EFT framework offers a unified approach for BSM searches at low energies and at colliders. At the momentum scale $\mu\sim v\equiv (\sqrt{2}G_F)^{-2}\approx 246.22$~GeV, one integrates out the BSM degrees of freedom and write down a Standard Model Effective Field Theory (SMEFT) that consists of the most general Lagrangian operating with SM fields SM gauge symmetries. Moving down in the energy scale to $\mu\sim 2$~GeV, one further integrates out the heavy gauge bosons, and transitions from SMEFT to the Low-Energy Effective Field Theory (LEFT). The leading order LEFT Lagrangian relevant to beta decays reads: \cite{Falkowski:2020pma}
\begin{align}
    \label{eq:TH_Lrweft}
{\mathcal L}  \supset 
{\mathcal L}^\mathrm{SM}
- \frac{V_{ud}}{v^2} \sum_i\Big[
\epsilon_i \,\bar{e}\, \Gamma_i\nu_L \,\bar u\,\Gamma_i d+\tilde\epsilon_i \,\bar e \,\Gamma_i\nu_R \,\bar u\,\Gamma_i d\Big]\,,
%
%&\left( 1 +  \eL \right) \ \bar{e}  \gamma_\mu  \nu_L  \cdot \bar{u}   \gamma^\mu  (1 - \gamma_5)  d \Big.
%~+~
%\teL  \,\bar{e}  \gamma_\mu \nu_R  \cdot \bar{u}   \gamma^\mu  (1 - \gamma_5)  d
%\nnl
%& +\eR   \,  \bar{e}  \gamma_\mu \nu_L
%\cdot \bar{u}   \gamma^\mu  (1 + \gamma_5)  d
%~+~
%\teR   \,  \bar{e}  \gamma_\mu    \nu_R
%\cdot \bar{u}   \gamma^\mu  (1 + \gamma_5)  d
%\nnl
%& +\frac{1}{4} \eT    \   \bar{e}   \sigma_{\mu \nu} \nu_L    \cdot  \bar{u}   \sigma^{\mu \nu} (1 - \gamma_5) d
%~+~
%\Big.
%\frac{1}{4} \teT      \   \bar{e}   \sigma_{\mu \nu}  \nu_R    \cdot  \bar{u}
 %\sigma^{\mu \nu} (1 + \gamma_5) d  
 %\nnl
%& + \eS  \, \bar{e}  \nu_L  \cdot  \bar{u} d
%~+~
%\teS  \, \bar{e}  \nu_R  \cdot  \bar{u} d
%\nnl
%&  
%- \eP  \,  \bar{e}  \nu_L  \cdot  \bar{u} \gamma_5 d
%~-~
%\teP  \,  \bar{e}  \nu_R  \cdot  \bar{u} \gamma_5 d
%\Big] + {\rm h.c.}~
%\label{eq:leff-lowE}
\end{align}
where $u$, $d$, $e$ are the up quark, down quark, and electron fields,
$\nu_{L,R} \equiv (1\pm\gamma_5)\nu/2$ are the left-handed and right-handed electron neutrino fields. The index $i=L,R,S,T,P$ counts structures that describe lefthanded, righthanded, scalar, tensor and pseudoscalar currents on the quark side, correspondingly. 
The Wilson coefficients $\epsilon_i$ and $\tilde{\epsilon}_j$ describe the coupling to left- and righthanded neutrinos, respectively, and vanish in absence of BSM. 

The momentum exchange in beta decays is far below the QCD scale, and the quark-level Lagrangian above is embedded in the nucleon-level Lee-Yang Lagrangian \cite{Lee:1956qn,Bhattacharya:2011qm,Cirigliano:2012ab,Gonzalez-Alonso:2018omy},
\begin{align}
    \label{eq:TH_Lleeyang}
\mathcal{L}_{\rm LY} &=
-  \bar{p}\gamma^\mu n 
\left( C_V^+ \bar{e} \gamma_\mu  \nu_L 
+ C_V^- \bar{e} \gamma_\mu \nu_R  \right) 
-  \bar{p}\gamma^\mu \gamma_5 n 
\left( C_A^+ \bar{e} \gamma_\mu  \nu_L 
-  C_A^- \bar{e} \gamma_\mu  \nu_R   \right)  
\nonumber\\
 &-
   \bar{p}n \left( 
C_S^+ \bar{e}  \nu_L + C_S^- \bar{e} \nu_R  \right)
- \frac{1}{2}\bar{p}\sigma^{\mu\nu} n 
 \left( C_T^+ \bar{e} \sigma_{\mu\nu} \nu_L 
 + C_T^- \bar{e} \sigma_{\mu\nu} \nu_R  \right) 
 \nonumber\\
 &+   \bar{p} \gamma_5 n  \left( C_P^+ \bar{e}\nu_L 
-  C_P^- \bar{e}\nu_R  \right)
+ \mathrm{h.c.}   
\end{align}
The quark- and nucleon level parameters are interrelated:  ~\cite{Gonzalez-Alonso:2018omy}
\begin{align}
    \label{eq:TH_LYtoRWEFT}
C_{V,A}^+ & =  \frac{V_{ud}}{v^2} \mathring{g}_{V,A} \sqrt{1 + \Delta_R^{V,A}} \big ( 1+ \epsilon_L \pm \epsilon_R \big ) , 
\quad 
C_{V,A}^-  =  \pm\frac{V_{ud}}{v^2} \mathring{g}_{V,A}  \sqrt{1 + \Delta_R^{V,A}}  \big ( \tilde \epsilon_L \pm \tilde \epsilon_R \big ) , 
\nnl 
%C_A^+ & = &  {V_{ud} \over v^2} \mathring{g}_A  \sqrt{1 + \Delta_R^A}  \big ( 1+ \epsilon_L - \epsilon_R \big ) , 
%\qquad 
%C_A^-  =  -{V_{ud} \over v^2} \mathring{g}_A  \sqrt{1 + \Delta_R^A}  \big ( \tilde \epsilon_L - \tilde \epsilon_R \big ) , 
%\nnl
C_i^+ & =  \frac{V_{ud}}{v^2} g_i \,\epsilon_i  \,, 
\quad 
C_i^-  =  \pm\frac{V_{ud}}{v^2} g_i \,\tilde \epsilon_i\,, \quad {\rm for}\;i=S,T,P\,,
%\nnl
%C_S^+ & = & {V_{ud} \over v^2} g_S \epsilon_S  , 
%\qquad 
%C_S^-  =  {V_{ud} \over  v^2} g_S \tilde \epsilon_S, 
%\nnl
%C_P^+ & = & {V_{ud} \over v^2} g_P \epsilon_P  , 
%\qquad 
%C_P^-  =  - {V_{ud} \over  v^2} g_P \tilde \epsilon_P. 
\end{align}
where the $+\,(-)$ sign should be taken for $V,S,T\,(A,P)$ couplings, respectively. 
We have already discussed the vector and axial nucleon charges $\mathring{g}_{V,A}$ in the previous sections; meanwhile, the FLAG'21 averages~\cite{Aoki:2021kgd} for the scalar and tensor charges read:  $g_S = 1.02(10)$ and $g_T = 0.989(34)$. 
Although the pseudoscalar charge is enhanced by the pion pole, $g_P=349(9)$~\cite{Gonzalez-Alonso:2013ura}, the kinematical suppression of the pseudoscalar contributions to the $\beta$ decay observables is still more significant.
The relations of Eq.~\eqref{eq:TH_LYtoRWEFT}, together with the known \cite{Jackson:1957auh,Jackson:1957zz} contribution of the Lee-Yang effective couplings to the total decay rate and correlation coefficients in Eq.~\eqref{eq:drate}, 
ultimately allow one to set constraints upon heavy new physics with beta decays.
\begin{figure}[h]
    \centering
\includegraphics[width=0.6\columnwidth]{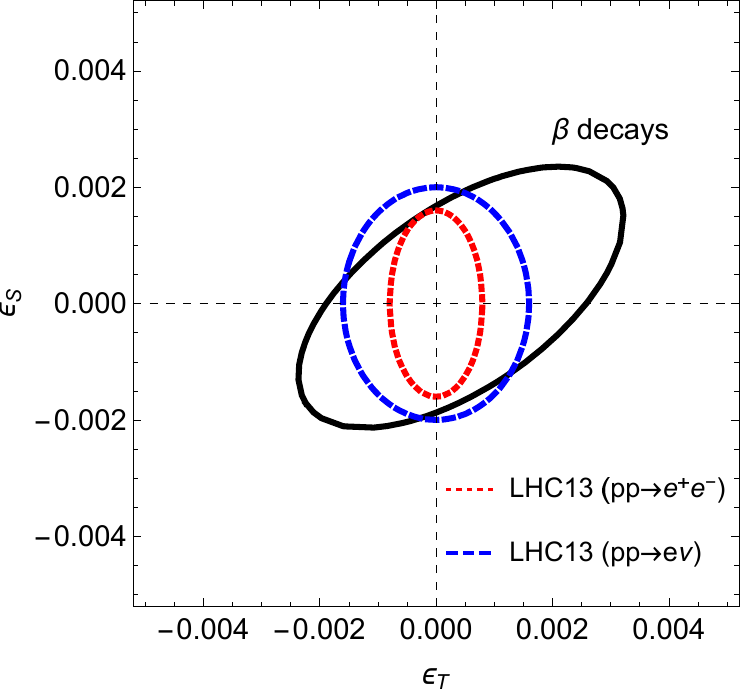}
    \caption{Constraints on scalar and tensor coupling constants from beta decays and LHC respectively. Figure reproduced from Ref.\cite{Falkowski:2020pma} with permission of the authors.}
    \label{fig:espST}
\end{figure}

Fig.~\ref{fig:espST} displays current limits on the Wilson coefficients $\epsilon_{S,T}$ 
obtained in Ref.~\cite{Falkowski:2021vdg} from beta decays and collider observables in the scenario of only two dimension-6 operators present at the TeV scale. Sensitivity to various non-standard interactions is very different at low and high energies because of the small electron and neutrino mass. $V,A$ interactions do not flip the lepton helicity and can be very efficiently probed at the LHC. Instead, $S,T$ involve a helicity flip, hence their contribution is suppressed as $m_e/E_e$ at high energy. Beta decays are free from this suppression, making their sensitivity to $\epsilon_{S,T}$ very competitive as compared to that of the LHC observables, as can be seen in Fig.~\ref{fig:espST}. Numerically, beta decay measurements translate into $10^{-3}$ level constraints, $\epsilon_S=0.0001(10)$, $\epsilon_T=0.0005(13)$ \cite{Falkowski:2021vdg}. 
The main source of constraints on $\epsilon_S$ remain superallowed nuclear decays, whereas those on $\epsilon_T$ mainly come from precise measurements of neutron decay. 
Note that since a very large number of SMEFT operators may contribute at the LHC energies, the respective limits on $\epsilon_{S,T}$ depicted in Fig.~\ref{fig:espST} were obtained in a simplified scenario switching all BSM operators but $S,T$ off. A complete SMEFT analysis is very complicated and will likely give much looser constraints on these two Wilson coefficients. 
%This limitation is not affecting low-energy beta decays where only a limited number of effective operators.
%In comparison, 8 TeV data from the LHC provide following bounds: 
%$|\epsilon_S|<0.005$, $|\epsilon_T|<0.0006$ \cite{CMS:2014fjm,ATLAS:2016gic}. With the 13 TeV upgrade and a higher integrated luminosity these limits will further improve. 
Recent global analyses of beta decays and LHC observables in terms of SMEFT operators can be found in \cite{Falkowski:2021vdg,Falkowski:2023klj} (see also Ref.\cite{Crivellin:2021bkd}). 
Heavy new physics explanations of the CKM unitarity deficit may require, e.g., lepton-flavor universality violation and quark-flavor dependent right-handed currents, and we refer the reader to  Ref.~\cite{Cirigliano:2022yyo} and references therein for a detailed discussion.

Furthermore, non-SMEFT new physics scenarios have been considered in the literature, mostly triggered by the stark beam-bottle disagreement for the neutron lifetime. If additional, nonstandard neutron decay channels would exist, the shorter lifetime obtained from counting the surviving neutrons in the bottle method could be accommodated with a smaller number of protons resulting from the neutron decay (and hence, a longer lifetime), as observed with the beam method  \cite{Fornal:2018eol}. We refer the reader to a dedicated review of dark and mirror neutron decay modes included in this issue, Ref.~\cite{Tan:2023mpj}. 
%%%%%%%%%%%%%%%%%%%%%%%%%%%%%%%%%%%%%%%%%%
\section{Conclusions \& Outlook}
\label{sec:conclusions}
In view of the currently accessible and ever improving experimental precision in measuring neutron lifetime and correlation coefficients, neutron decay is one of our best avenues to test the SM and its extensions at the relative level of few parts in $10^4$. Future measurements are expected to bring $V_{ud}$ extracted from neutron decay to $0.01-0.02\%$. 
To empower this program, SM RC to similar or better precision are necessary. In this review we provided an in-depth overview of recent developments in this area. In particular, the recent factor 2 reduction of the uncertainty of the inner radiative corrections $\Delta_R^{V,A}$ achieved in the novel dispersive approach combined with inputs from experimental data, perturbative QCD, chiral EFT, lattice QCD and Regge theory ensures that the aforementioned future experimental improvements will not be blurred by theory uncertainties. This situation is different from that in the superallowed nuclear decays where, following the recent scrutiny of nuclear structure effects, the uncertainty of the latter is by far the limiting factor. This situation marks the shift of paradigm in beta decays: after decades of the domination of the superallowed nuclear decays as the primary source of $V_{ud}$, for the first time the neutron comes center stage on par with the latter.

At the $10^{-4}$ level of accuracy, beta decays of free and bound neutrons allow to set stringent constraints on possible SM extensions. In this perspective, having both neutron and nuclear decays at a similar level of precision fosters sensitive tests of nonstandard scalar and tensor currents. The two beta decay channels show different sensitivity to scalar and tensor BSM contributions thus being complementary to each other. Importantly, the impact of high-precision neutron and nuclear beta decay as a sensitive avenue to searches for new physics will persist even into the high-luminosity LHC era. 

%%%%%%%%%%%%%%%%%%%%%%%%%%%%%%%%%%%%%%%%%%
%\section{Patents}

%This section is not mandatory, but may be added if there are patents resulting from the work reported in this manuscript.

%%%%%%%%%%%%%%%%%%%%%%%%%%%%%%%%%%%%%%%%%%
\vspace{6pt}

{\bf Acknowledgements}: 
{The authors thank Vincenzo Cirigliano for useful discussions. The work of M.G. is supported in part by EU Horizon 2020 research and innovation programme, STRONG-2020 project
under grant agreement No 824093, and by the Deutsche Forschungsgemeinschaft (DFG) under the grant agreement GO 2604/3-1. The work of C.Y.S. is supported in
part by the U.S. Department of Energy (DOE), Office of Science, Office of Nuclear Physics, under the FRIB Theory Alliance award DE-SC0013617, and by the DOE grant DE-FG02-97ER41014. We acknowledge partial
support from the DOE Topical Collaboration ``Nuclear Theory for New Physics'', award No.
DE-SC0023663. }

%\conflictsofinterest{The authors declare no conflict of interest.} 

%%%%%%%%%%%%%%%%%%%%%%%%%%%%%%%%%%%%%%%%%%
%% Optional
%\sampleavailability{Samples of the compounds ... are available from the authors.}

%% Only for journal Encyclopedia
%\entrylink{The Link to this entry published on the encyclopedia platform.}

%\abbreviations{Abbreviations}{
%The following abbreviations are used in this manuscript:\\

%\noindent 
%\begin{tabular}{@{}ll}
%MDPI & Multidisciplinary Digital Publishing Institute\\
%DOAJ & Directory of open access journals\\
%TLA & Three letter acronym\\
%LD & Linear dichroism
%\end{tabular}
%}

%%%%%%%%%%%%%%%%%%%%%%%%%%%%%%%%%%%%%%%%%%
%% Optional
%\appendixtitles{yes} % Leave argument "no" if all appendix headings stay EMPTY (then no dot is printed after "Appendix A"). If the appendix sections contain a heading then change the argument to "yes".
%\appendixstart
\begin{appendix}

%\subsection[\appendixname~\thesubsection]{}
%The appendix is an optional section that can contain details and data supplemental to the main text---for example, explanations of experimental details that would disrupt the flow of the main text but nonetheless remain crucial to understanding and reproducing the research shown; figures of replicates for experiments of which representative data are shown in the main text can be added here if brief, or as Supplementary Data. Mathematical proofs of results not central to the paper can be added as an appendix.

\section{Kinematics}
\label{sec:kinematics}
\begin{figure}[h]
    \centering
\includegraphics[width=0.67\columnwidth]{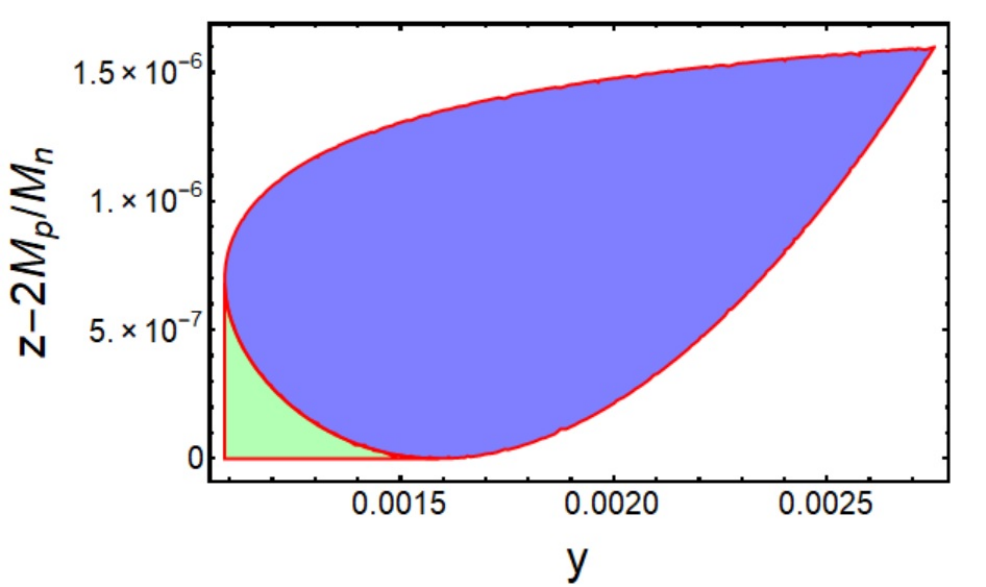}
    \caption{Illustration of the $\mathcal{D}_3$ region (blue) and $\mathcal{D}_{4-3}$ region (green) in free neutron decay. }
    \label{fig:Dalitz}
\end{figure}
The process of primary interest is $n\rightarrow p + e + \bar{\nu}_e$, accompanied by the 
experimentally indistinguishable inclusive process $n\rightarrow p + e + \bar{\nu}_e + N\gamma$ with $N\geq 0$ photons. To match the current experimental precision only $N=0,1$ are needed. If massless particles ($\bar{\nu}_e$ and $\gamma$) are unobserved, there are only three independent Lorentz-invariant kinematic variables:
\begin{equation}
x\equiv \frac{P^2}{M_n^2}~,~y\equiv\frac{2p_n\cdot p_e}{M_n^2}~,~z\equiv \frac{2p_n\cdot p_p}{M_n^2}~,
\end{equation}
with $P\equiv p_n-p_p-p_e$ the sum of the momenta of the unobserved massless particles, and $r_p\equiv M_p^2/M_n^2$ and $r_e\equiv m_e^2/M_n^2$. For $N=0$ the three-body kinematics $\mathcal{D}_3$ is defined by
\begin{eqnarray}
a(y)-b(y) < z < a(y)+b(y) &,& 2\sqrt{r_e}<y<1+r_e-r_p  \nonumber\\
a(y)\equiv\frac{(2-y)(1+r_p+r_e-y)}{2(1+r_e-y)}&,&b(y)\equiv\frac{\sqrt{y^2-4r_e}(1+r_e-r_p-y)}{2(1+r_e-y)}
\end{eqnarray}
or equivalently,
\begin{eqnarray}
c(z)-d(z) < y < c(z)+d(z) &,& 2\sqrt{r_p}<z<1+r_p-r_e  \nonumber\\
c(z)\equiv\frac{(2-z)(1+r_e+r_p-z)}{2(1+r_p-z)}&,&d(z)\equiv\frac{\sqrt{z^2-4r_p}(1+r_p-r_e-z)}{2(1+r_p-z)}~.
\end{eqnarray}
Notice that $x=0$ for $N=0$ assuming massless neutrino. 
For $N=1$, the four-body kinematics $\mathcal{D}_4$ consists of two regions depicted in Fig.\ref{fig:Dalitz}:
\begin{equation}
0<x<\alpha_+(y,z)~,~(y,z)\in\mathcal{D}_3~~\text{and}~~\alpha_-(y,z)<x<\alpha_+(y,z)~,~(y,z)\in\mathcal{D}_{4-3}~,
\end{equation}
where $\alpha_\pm (y,z)\equiv 1-y-z+r_p+r_e+yz/2\pm(1/2)\sqrt{y^2-4r_e}\sqrt{z^2-4r_p}$. There are also two equivalent way to represent $\mathcal{D}_{4-3}$:
\begin{align}
   & 2\sqrt{r_e}<y<c(z)-d(z)\,,\quad2\sqrt{r_p}<z<1-\sqrt{r_e}+\frac{r_p}{1-\sqrt{r_e}}\,.\quad\mathrm{or}\nonumber\\
&2\sqrt{r_p}<z<a(y)-b(y)\,,\quad2\sqrt{r_e}<y<1-\sqrt{r_p}+\frac{r_e}{1-\sqrt{r_p}}~.
\end{align}

In the existing literature, the differential decay rate is always expressed in terms of $\vec{p}_\nu$ rather than $\vec{P}$, despite the fact that it is the latter that is an experimental observable. In this case, an extra step to theoretically subtract out the observed photon momentum is necessary. The failure to do so could lead to an error in the extraction of $\lambda$ from certain correlation coefficients in the differential rate, e.g. the coefficients ``$a$'' and ``$B$'' in Sec.\ref{sec:diffrate}. This was pointed out in Refs.\cite{Gluck:1989sf,Gluck:1992qy,Gluck:1997km,Hayen:2020nej,Gluck:2022ogz}.
\end{appendix}

%\section[\appendixname~\thesection]{}
%All appendix sections must be cited in the main text. In the appendices, Figures, Tables, etc. should be labeled, starting with ``A''---e.g., Figure A1, Figure A2, etc.

%%%%%%%%%%%%%%%%%%%%%%%%%%%%%%%%%%%%%%%%%%
%\begin{adjustwidth}{-\extralength}{0cm}
%\printendnotes[custom] % Un-comment to print a list of endnotes

%\reftitle{References}

% Please provide either the correct journal abbreviation (e.g. according to the “List of Title Word Abbreviations” http://www.issn.org/services/online-services/access-to-the-ltwa/) or the full name of the journal.
% Citations and References in Supplementary files are permitted provided that they also appear in the reference list here. 

%=====================================
% References, variant A: external bibliography
%=====================================
%\bibliography{your_external_BibTeX_file}

%=====================================
% References, variant B: internal bibliography
%=====================================
%\externalbibliography{yes}
%\bibliography{neutron_ref}
\bibliography{main.bbl}

% If authors have biography, please use the format below
%\section*{Short Biography of Authors}
%\bio
%{\raisebox{-0.35cm}{\includegraphics[width=3.5cm,height=5.3cm,clip,keepaspectratio]{Definitions/author1.pdf}}}
%{\textbf{Firstname Lastname} Biography of first author}
%
%\bio
%{\raisebox{-0.35cm}{\includegraphics[width=3.5cm,height=5.3cm,clip,keepaspectratio]{Definitions/author2.jpg}}}
%{\textbf{Firstname Lastname} Biography of second author}

% For the MDPI journals use author-date citation, please follow the formatting guidelines on http://www.mdpi.com/authors/references
% To cite two works by the same author: \citeauthor{ref-journal-1a} (\citeyear{ref-journal-1a}, \citeyear{ref-journal-1b}). This produces: Whittaker (1967, 1975)
% To cite two works by the same author with specific pages: \citeauthor{ref-journal-3a} (\citeyear{ref-journal-3a}, p. 328; \citeyear{ref-journal-3b}, p.475). This produces: Wong (1999, p. 328; 2000, p. 475)

%%%%%%%%%%%%%%%%%%%%%%%%%%%%%%%%%%%%%%%%%%
%% for journal Sci
%\reviewreports{\\
%Reviewer 1 comments and authors’ response\\
%Reviewer 2 comments and authors’ response\\
%Reviewer 3 comments and authors’ response
%}
%%%%%%%%%%%%%%%%%%%%%%%%%%%%%%%%%%%%%%%%%%
%\PublishersNote{}

\end{document}